\documentclass[12pt,a4paper]{article}

\usepackage{authblk}
\usepackage{amsmath,amssymb}
\usepackage{graphicx}
\usepackage{amsfonts}
\usepackage{epsfig}
\usepackage{times,pstricks,pst-node}

\newcommand{\im}{\mathrm i}

\newcommand{\tr}{\operatorname{Tr}}
\newcommand{\llg}{\operatorname{log}}

\newcommand{\ket}[1]{\left|#1\right\rangle}      
\newcommand{\bra}[1]{\left\langle #1\right|}     
\newcommand{\eq}{\begin{equation}}
\newcommand{\en}{\end{equation}}
\newcommand{\bear}{\begin{eqnarray}}
\newcommand{\ear}{\end{eqnarray}}

\title{Correlation functions of the integrable spin-$s$ chain}

\author{G.A.P. Ribeiro\footnote{E-mail: pavan@df.ufscar.br} \ and A. Kl\"umper\footnote{E-mail: kluemper@physik.uni-wuppertal.de}}
\affil{$^{*}$ Departamento de F\'{i}sica, Universidade Federal de S\~ao Carlos \\ S\~ao Carlos, SP 13565-905, Brazil
\\
$^{\dagger}$ Theoretische Physik, Bergische Universit\"at Wuppertal, \\ 42097
Wuppertal, Germany}
\date{}

\begin{document}
\maketitle
\begin{center}
{\it In honor of Rodney Baxter's 75th birthday}
\end{center}

\begin{abstract}
We study the correlation functions of $su(2)$ invariant spin-$s$ chains in the
thermodynamic limit. We derive non-linear integral equations for an auxiliary
correlation function $\omega$ for any spin $s$ and finite temperature $T$.
For the spin-$3/2$ chain for arbitrary temperature and zero magnetic field we
obtain algebraic expressions for the reduced density matrix of two-sites. In
the zero temperature limit, the density matrix elements are evaluated
analytically and appear to be given in terms of Riemann's zeta function values
of even and odd arguments.
\end{abstract}

\thispagestyle{empty}
\newpage

\pagestyle{plain}
\pagenumbering{arabic}

\section{Introduction}

The static correlation functions of quantum integrable models, 
most notably the spin-$1/2$ Heisenberg model, have been extensively studied
over the years. The first results were presented in terms of multiple
integrals \cite{JMMN92,JM96,KMT00,GAS05}. In the course of an explicit
evaluation of these integrals a factorization in terms of sums over products
of single integrals was found \cite{BOKO01}. These results were extended to
finite temperature in the thermodynamic limit and to zero temperature and
finite chains \cite{BGKS06,DGHK07}. In this context, a hidden Grassmann
structure was identified \cite{BJMST06}, which made possible to prove the
complete factorization of the correlation functions under general
conditions \cite{JMS08}.

The multiple integral representation of the spin-$1/2$ chain was successfully
used to obtain, from first principles, the long distance asymptotic behavior
of the correlation functions, which confirmed at leading order the conformal
field theory predictions \cite{KKMST09}. Moreover, explicit values for the
short range correlations were systematically obtained by direct evaluation of
multiple integrals \cite{BOKO01} and later by solving recurrently functional
equations obtained from the quantum Knizhnik Zamolodchikov
equation \cite{BOKO03,BST05,SST05}. These correlation functions are remarkably
given in terms of combinations of the zeta function with odd integer
arguments \cite{BOKO01}.

Nevertheless, one still lacks a better understanding of the correlation
properties of models based on higher rank algebras and higher-spin
representations of the $su(2)$ algebra. For the latter case, the high-spin
generalizations, there exist multiple integral representations of the
correlation functions of the ground
state \cite{BoWe94,Idzumi94,Kitanine01,DeMa10} and for finite
temperature \cite{GSS10}. However, the expressions are very intricate for
explicit evaluations. On the other hand, the explicit computation of the short
range correlations of the integrable spin-$1$ chain \cite{Babujian82,TAK} has
been done relatively recently \cite{KNS013} by means of the solution of
discrete functional equations \cite{AuKl12}. Surprisingly, the correlation
functions for two-sites and three-sites are given in terms of powers of
$\pi^2$ or alternatively in terms of the zeta function of only even integer
arguments. 

In this work, we are interested in applying the approach of discrete
functional equations to evaluate the first non-trivial correlation function,
for two-sites, in case of the integrable spin-$3/2$ chain. We use a systematic
approach to tackle the high-spin chains in analogy to \cite{KNS013}.  This
approach uses an auxiliary spin-$1/2$ nearest neighbour correlation function
$\omega$ and fusion principles.  In making it systematically work for spin
$s>1$ we found an important auxiliary condition of the function $\omega$ which
did not appear yet for $s=1$. Our procedure yields the solution for the
two-sites correlation function at finite and zero temperature, proving that
the approach is viable for obtaining explicit results. Amazingly, our explicit
results are in contrast to a naive conjecture based on the spin-$1/2$ and $1$
results, that the correlations for half odd integer (integer) spin chains
may be given just in terms of zeta function values with odd (even) integer
arguments. In fact, our results already for the two-sites of the spin-$3/2$
chain are a combination of $\pi^2$ and $\log2$, which can be seen as zeta
function values of even and odd arguments.

This paper is organized as follows. In section \ref{INTEGRA}, we outline the
integrable Hamiltonians and the associated integrable structure. In
section \ref{density}, we introduce the physical density matrix and its
functional equations. The physical properties of the model are given in terms
of non-linear integral equations in section \ref{NLIE}. In
section \ref{2siteDM}, we present the two-site density matrix for the spin-$s$
chains for $s$ up to $3/2$. Our conclusions are given in
section \ref{CONCLUSION}.

\section{The model}\label{INTEGRA}

The Hamiltonians of the integrable spin-$s$ generalization of the Heisenberg
model for $s=1/2,1$ and $3/2$ are given by
\bear
{\cal H}^{[1]}&=&J\sum_{i=1}^{L}[\frac{1}{4}+\vec{S}_i \cdot \vec{S}_{i+1}], \nonumber\\
{\cal H}^{[2]}&=&\frac{J}{4}\sum_{i=1}^{L} [3+ \vec{S}_i \cdot \vec{S}_{i+1} - (\vec{S}_i \cdot \vec{S}_{i+1})^2], \\
{\cal H}^{[3]}&=&\frac{J}{532}\sum_{i=1}^{L}
[234-27\vec{S}_i \cdot \vec{S}_{i+1} 
 + 8(\vec{S}_i \cdot \vec{S}_{i+1})^2+ 16(\vec{S}_i \cdot \vec{S}_{i+1})^3],\nonumber
\ear
where $L$ is the number of sites and
$\vec{S}_{i}=(\hat{S}_{i}^{x},\hat{S}_{i}^{y},\hat{S}_{i}^{z})$ are the $su(2)$ generators and $J$ is the exchange constant. Here, it is worth to
recall that the Hamiltonian ${\cal H}^{[2s]}$ is obtained from the logarithmic
derivative of the row-to-row transfer matrix 
$t^{[2j]}(\lambda)=\tr_{\cal A}{[ R_{{\cal A},L}^{[2j,2s]}(\lambda)\dots
R_{{\cal A}, 1}^{[2j,2s]}(\lambda)]}$, which results into
\eq
{\cal H}^{[2s]}=\im
J \frac{d}{d\lambda}\log{(t^{[2s]}(\lambda))}\Big|_{\lambda=0} 
=J \sum_{i=1}^L h_{i,i+1}, \qquad 
h_{i,i+1}=\im \frac{d}{d\lambda} \check{R}^{[2s,2s]}_{i,i+1}\Big|_{\lambda=0},
\en
where $\check{R}^{[2s,2s]}_{12} = P_{12}^{[2s]} R^{[2s,2s]}_{12}(\lambda)$, $P_{12}^{[2s]}$ is the permutation operator and $L$ is the number of sites. The $R$-matrix $R^{[2j,2s]}_{12}(\lambda)$ is a solution of the Yang-Baxter equation
\eq
R_{12}^{[2s_1,2s_2]} (\lambda-\mu) R_{13}^{[2s_1,2s_3]} (\lambda) R_{23}^{[2s_2,2s_3]} (\mu) =R_{23}^{[2s_2,2s_3]}(\mu) R_{13}^{[2s_1,2s_3]}(\lambda)  R_{12}^{[2s_1,2s_2]} (\lambda-\mu),
\label{yang-baxter}
\en
which has the properties of regularity and unitarity as given below,
 \bear
 R_{12}^{[2s,2s]}(0)=P_{12}^{[2s]}, \\
 R_{12}^{[2s,2s]} (\lambda) R_{12}^{[2s,2s]} (-\lambda) = \mbox{Id}.
 \ear

The operator $R_{12}^{[2s_1,2s_2]}(\lambda)$ is a rational solution of
the Yang-Baxter equation and is obtained, for instance, by the fusion
process \cite{KULISH},
\eq
R_{12}^{[2s_1,2s_2]}(\lambda)= \sum_{l=|s_1-s_2|}^{s_1+s_2}
f_{l}(\lambda) \check{P}_{l},
\label{Loperator}
\en
where $ f_{l}(\lambda)=\prod_{k=l+1}^{s_1+s_2}
(\frac{\lambda- \im 2 k}{\lambda+ \im 2 k})$ and $\check{P}_l$ is the
usual $su(2)_l$ projector
\eq
\check{P}_{l}=\prod_{\stackrel{k=|s_1-s_2|}{k \neq l}}^{s_1+s_2}
\frac{\vec{S}_{1}\cdot \vec{S}_{2}-x_{k}}{x_{l}-x_{k}},
\label{su2projector}
\en
with $x_{k}=\frac{1}{2}\left[k(k+1)-s_1(s_1+1)-s_2(s_2+1)\right]$.

More generally and compactly, the Hamiltonians can be written as
\eq
{\cal H}^{[2s]}=\frac{J}{2}\sum_{i=1}^{L} Q_{2s}(\vec{S}_i \cdot \vec{S}_{i+1}),
\label{Hamiltonian}
\en
where
\eq
Q_{2s}(x)=\sum_{i=0}^{2s}\left[2\Psi(i+1)-\Psi(1) -\Psi(2s+1)\right] \prod_{\stackrel{k=0}{k\neq i}}^{2s} \frac{x-x_k}{x_i-x_k},
\en
and $\Psi(x)$ is the digamma function.

\section{Density matrix	}\label{density}

The formalism to calculate the thermal correlation functions of quantum
integrable Hamiltonians derived from the row-to-row transfer matrix was
established in \cite{GoKlSe04}. This formalism can be understood as a lattice
path integral formulation and the evaluation of physical properties by use
of the quantum transfer matrix (QTM) \cite{MSUZUKI}. The quantum chain at
finite temperature is mapped by an integrable version of Trotter-Suzuki
relations onto a two-dimensional staggered vertex model on a $L\times N$
lattice. The statistical operator $\rho_L$ of the finite temperature quantum
chain is then given in terms of row-to-row transfer matrices
\eq
\rho_L=e^{-\beta {\cal H}^{(s)}}= \lim_{N\rightarrow \infty} \rho_{N,L}, \qquad \rho_{N,L}= \left[ t^{(2s)}(-\beta/N)\bar{t}^{(2s)}(-\beta/N) \right]^{N/2},
\en
where ${t}^{(2s)}(\lambda)$ and $\bar{t}^{(2s)}(\lambda)$ are the usual and
the adjoint transfer matrices. For more details see for
instance \cite{GoKlSe04}. For convenience, the $\rho_{N,L}$ is rewritten along
the quantum direction in terms of the column-to-column monodromy matrix
${\cal T}^{[2s]}(\lambda)$,
\eq
\rho_{N,L}=\tr_{1\cdots N}\left[{\cal T}^{[2s]}_1(0)\cdots {\cal T}^{[2s]}_L(0)\right],
\en
where the trace over spaces $1,..., N$ takes account of periodic boundary
conditions in horizontal direction and
\eq
{\cal T}^{[2s]}_i(\lambda)=R^{[2s,2s]}_{i,N}(\lambda+\im u)R^{[2s,2s]t_{N-1}}_{N-1,i}(\im u-\lambda)\cdots R^{[2s,2s]}_{i,2}(\lambda+\im u)R^{[2s,2s]t_{1}}_{1,i}(\im u-\lambda), 
\en
where $u=-J\beta/N$ and at every second $R$-matrix the $t_k$ denotes
transposition in the $k$th space (which actually is the first space of the concerned $R$-matrix).

Taking the trace over the quantum space $i$ ($\in\{1,...,L\}$) realizes
periodic boundary conditions in vertical direction and results in
\eq
T^{[2s]}(\lambda)=\tr_i\left[{\cal T}^{[2s]}_i(\lambda)\right],
\en
which is called the quantum transfer matrix. Note that it acts in the
auxiliary space of states along the vertical direction. Thanks to the
Yang-Baxter equation (\ref{yang-baxter}), this transfer matrix forms a family
of commuting operators $[T^{[2s]}(\lambda),T^{[2s]}(\mu)]=0$, which implies
the quantum transfer matrix eigenvectors $\ket{\Phi_m}$ do not depend on the
spectral parameter.  This matrix can be diagonalized by Bethe ansatz
techniques \cite{AK93,JSUZUKI99} and quite generally it has a non-degenerate
largest eigenvalue $\Lambda_0^{[2s]}(0)$ split from the rest of the spectrum
by a gap \cite{MSUZUKI,JSUZUKI90}.

The thermodynamic properties of the quantum Hamiltonian are obtained via the
partition function $Z_L=\tr_{1\cdots L}\left[e^{-\beta {\cal
H}^{(s)}}\right]=\lim_{N\rightarrow \infty}\tr_{1\cdots
L}\left[\rho_{N,L} \right]$. As \break $\tr_{1\cdots
L}\left[\rho_{N,L} \right]=\tr_{1\cdots N}\left[T^{[2s]}(0)\right]^L$, the
free energy in the thermodynamic limit is given by the leading eigenvalue
$\Lambda^{[2s]}_0$ of the quantum transfer matrix $T^{[2s]}$
\eq
f^{[2s]}=-\frac{1}{\beta}\lim_{N,L\rightarrow\infty}\frac{1}{L} \log Z_L = -\frac{1}{\beta}\lim_{N\rightarrow\infty}\log\Lambda^{[2s]}_0(0).
\label{free-energy}
\en
The leading eigenvalue, especially in the Trotter limit $N\to\infty$, is
obtained \cite{AK93,JSUZUKI99} in terms of the solution of a set of non-linear
integral equations (see section \ref{NLIE}).

The set of all thermal correlation functions with finite separation of the
local operators is encoded in the reduced density matrix
on a finite chain segment of some length $n$,
\bear
D_{[1,n]}^{[2s]}&=& \lim_{L\rightarrow\infty} \frac{\tr_{n+1\cdots L}\left[\rho_L \right]}{Z_L}=\lim_{N,L\rightarrow\infty}\frac{\tr_{n+1\cdots L}\left[\rho_{N,L} \right]}{\tr_{1\cdots L}\left[\rho_{N,L} \right]},
\label{density1}
\ear
Introducing in (\ref{density1}) a complete set of eigenstates of the quantum
transfer matrix we obtain \cite{GoKlSe04},
\bear
D_{[1,n]}^{[2s]}&=& \lim_{N,L\rightarrow\infty}\frac{\tr_{1\cdots N}\tr_{n+1\cdots L}\left[{\cal T}^{[2s]}_1(0)\cdots {\cal T}^{[2s]}_L(0)\right]}{\tr_{1\cdots N}\tr_{1\cdots L}\left[{\cal T}^{[2s]}_1(0)\cdots {\cal T}^{[2s]}_L(0)\right]}, \nonumber\\
&=& \lim_{N,L\rightarrow\infty}\frac{\tr_{1\cdots N}\left[{\cal T}^{[2s]}_1(0)\cdots {\cal T}^{[2s]}_n(0) (T^{[2s]}(0))^{L-n} \right]}{\tr_{1\cdots N} (T^{[2s]}(0))^L}, \nonumber\\
&=& \lim_{N,L\rightarrow\infty} \frac{ \sum_m (\Lambda_m^{[2s]}(0))^{L-n} \bra{\Phi_m}{\cal T}^{[2s]}_1(0)\cdots {\cal T}^{[2s]}_n(0) \ket{\Phi_m}}{\sum_m  (\Lambda_m^{[2s]}(0))^L}, \nonumber\\
&=& \lim_{N\rightarrow\infty} \frac{  \bra{\Phi_0}{\cal T}^{[2s]}_1(0)\cdots {\cal T}^{[2s]}_n(0) \ket{\Phi_0}}{ (\Lambda_0^{[2s]}(0))^n},
\ear
where in the end the thermodynamic limit was taken by dropping the sub-dominant
states with $m>0$. This implies that all the static correlation functions at
finite temperature are determined by the dominant eigenvector $\ket{\Phi_0}$.

For technical reasons, it is convenient to introduce mutually distinct
spectral parameters $(\xi_1,\cdots,\xi_n)$ along the vertical lines 
corresponding to the quantum states $1,..., n$
in the expression for the reduced density matrix, such that,
\eq
D_n^{[2s]}(\xi_1,\cdots,\xi_n)=\frac{  \bra{\Phi_0}{\cal T}^{[2s]}_1(\xi_1)\cdots {\cal T}^{[2s]}_n(\xi_n) \ket{\Phi_0}}{ \Lambda_0^{[2s]}(\xi_1)\cdots \Lambda_0^{[2s]}(\xi_n) },
\label{i-dm}
\en
which is called the inhomogeneous density matrix with finite Trotter
number. This object enjoys interesting relations as a function of the variables
$(\xi_1,\cdots,\xi_n)$. These relations will be used for the explicit
computations. The final physically interesting result is obtained by taking
the homogeneous limit
\eq
D^{[2s]}_{[1,n]}=\lim_{N\rightarrow\infty} \lim_{\xi_1,\cdots,\xi_n\rightarrow 0} D_n^{[2s]}(\xi_1,\cdots,\xi_n).
\en

\subsection{Discrete functional equations}

The general framework to obtain discrete functional equations for the reduced
density matrix of integrable models on semi-infinite ($N \times \infty$)
lattices was established in \cite{AuKl12,KNS013}.

\begin{figure}[h]
\begin{center}
\includegraphics[width=\linewidth]{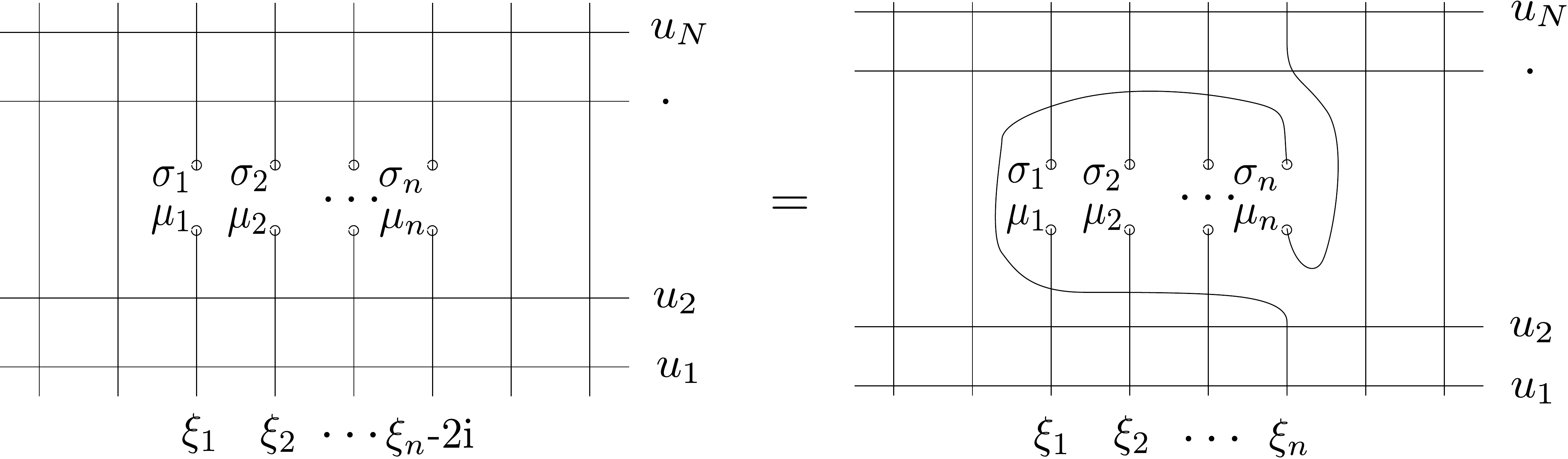}
\caption{Graphical illustration of the functional equation (\ref{qKZ}).}
\label{pic-qKZ}
\end{center}
\end{figure}

The general idea consists in starting with a slightly more general density
matrix, where $N$ many different spectral parameters $u_i$ are introduced on
the $N$ many horizontal lines. The derivation of the discrete functional
equations relies on the standard unitarity, regularity and crossing symmetry
of the $R$-matrix, which allows to perform the lattice manipulations described
in \cite{AuKl12}. The lattice surgery is allowed when one of the spectral
parameters $\xi_1,\cdots,\xi_n$, let us say the last $\xi_n$, is set identical to
any of the spectral parameters along the horizontal lines. This results in a
mapping of $D_n^{[2s]}(\xi_1,\cdots,\xi_n)$ to
$D_n^{[2s]}(\xi_1,\cdots,\xi_n-2\im)$ by a linear map $A$ acting on the space
of reduced density matrices. The resulting equations are the discrete quantum
Knizhnik-Zamolodchikov (qKZ) functional equations, 
\eq
D_n^{[2s]}(\xi_1,\cdots,\xi_n-2\im)=A_n^{[2s]}(\xi_1,\cdots,\xi_n)[D_n^{[2s]}(\xi_1,\cdots,\xi_n)],
\label{qKZ}
\en
where the linear operator $A_n^{[2s]}$ can be written as \cite{AuKl12},
\bear
A_n^{[2s]}(\xi_1,\cdots,\xi_n)[B]&:=&\tr_{V_n}\Big[R_{1,2}^{[2s,2s]}(-\xi_{1,n}) \cdots R_{n-1,n}^{[2s,2s]}(-\xi_{n-1,n})(P_{s})_{n,n+1} \nonumber \\
  &\times &(B\otimes I_{n+1}) R_{n-1,n}^{[2s,2s]}(\xi_{n-1,n})\cdots R_{1,2}^{[2s,2s]}(\xi_{1,n})\Big].
  \label{A-qKZ}
\ear
Here we use the short-hand $\xi_{i,j}=\xi_i-\xi_j$ and $P_{s}$ is a (not
normalized) projector onto the two-site singlet. The graphical depiction of of
the functional equation (\ref{qKZ}) is given in Figure \ref{pic-qKZ}.
Moreover, we have the asymptotic condition,
\eq
\lim_{\xi_n\rightarrow\infty} D_n^{[2s]}(\xi_1,\cdots,\xi_n)=D_{n-1}^{[2s]}(\xi_1,\cdots,\xi_{n-1})\otimes \mbox{Id}.
\en
Besides, the under-determinacy of the functional equations for the density matrix as a function of a single variable is resolved by exploiting the full dependence on all variables, fusion and using the intertwining symmetry relations
\bear
R_{k,k+1}^{[2s,2s]}(\xi_k-\xi_{k+1})D_n^{[2s]}(\xi_1,\cdots,\xi_k,\xi_{k+1},\cdots,\xi_n)(R^{[2s,2s]}_{k,k+1}(\xi_k-\xi_{k+1}))^{-1} =  \nonumber\\ 
= D_n^{[2s]}(\xi_1,\cdots,\xi_{k+1},\xi_{k},\cdots,\xi_n).
\ear
We avoid the explicit application of the intertwining symmetry by an approach
already used in \cite{KNS013}. By viewing each of the spin-$s$ horizontal
lines as a result of $2s$ many spin-$1/2$ lines in a suitable limit, it was
argued in \cite{KNS013} that the famous factorization expressions of arbitrary
spin-$1/2$ correlators in terms of nearest-neighbour spin-$1/2$ correlators
$\omega$ hold literally as in \cite{BJMST04a,BST05,SST05}. 
Then in principle, by taking the fusion
of $2s$ many vertical spin-$1/2$ lines to produce vertical spin-$s$ lines, the
results for the spin-$s$ correlators can be derived. We do
not follow this procedure literally. Instead of using the explicit
coefficients of multilinear expressions in $\omega$ as worked out for the
spin-$1/2$ case by \cite{BST05,SST05}, we use their principle of the
analytical structure of the coefficients and apply this directly to the high
spin case.

In the next section, the function $\omega$ will be calculated for arbitrary
spin $s$ and $T>0$. The computation of the coefficients of the multilinear
expressions in $\omega$ for the spin $s$ case will be summarized in section
\ref{2siteDM}.

\section{Bethe ansatz and non-linear integral equations}\label{NLIE}

In order to evaluate the free energy (\ref{free-energy}) and the density
matrix (\ref{i-dm}) knowledge of the leading eigenvalue
$\Lambda_0^{[2s]}(\lambda)$ and the associated eigenvector $\ket{\Phi_0}$ is
required.

As we are going to use fusion relations for the QTM, it is convenient to
consider additional staggered monodromy matrices with arbitrary spin-$j$ in
the vertical space, $N$-many spin-$s$ spaces and additional
two spin-$1/2$ spaces along the horizontal direction as given by
\bear
{\cal T}^{[2j]}_i(\lambda)&=&R^{[2j,1]}_{i,N+2}(\lambda-\mu)R^{[1,2j]t_{N+1}}_{N+1,i}(\mu+\delta-\lambda)  \label{monodromy-mod}\\
&\times&R^{[2j,2s]}_{i,N}(\lambda+\im u)R^{[2s,2j]t_{N-1}}_{N-1,i}(\im u-\lambda)\cdots R^{[2j,2s]}_{i,2}(\lambda+\im u)R^{[2s,2j]t_1}_{1,i}(\im u-\lambda), \nonumber
\ear
where in the case of $j=1/2$ the two spin-$1/2$ spaces will be used for
generating the auxiliary spin-1/2 nearest neighbour correlator
$\omega$, see \cite{AuKl12,KNS013} and the construction
below.

The trace of the monodromy matrix (\ref{monodromy-mod}) results in the
generalized quantum transfer matrix,
\eq
T^{[2j]}(\lambda)=\tr_{i}\left[{\cal T}^{[2j]}_i(\lambda)\right].
\label{QTMstag}
\en
The transfer matrix is known to satisfy the fusion hierarchy, which
reads \cite{KULISH,JSUZUKI99}
\begin{align}
&T^{[2j]}(\lambda)T^{[1]}(\lambda+\im(2j+1))=
T^{[2j+1]}(\lambda+\im ) +
\chi(\lambda+\im (2 j))T^{[2j-1]}(\lambda-\im), \nonumber\\
&T^{[0]}(\lambda)= \mbox{Id}, ~~ j=1/2,1,3/2\dots 
\label{fusion}
\end{align}
where $\chi(\lambda)=\frac{\phi_{-}(\lambda+\im( 2 s))\phi_{+}(\lambda-\im(2 s))}{\phi_{-}(\lambda-\im( 2 s))\phi_{+}(\lambda+\im(2 s))} \frac{\varphi_{-}(\lambda+\im)\varphi_{+}(\lambda-\im)}{\varphi_{-}(\lambda-\im)\varphi_{+}(\lambda+\im)}$ is the quantum determinant and
$\phi_{\pm}(\lambda)=(\lambda \pm \im u)^{N/2}$ and $\varphi_+(\lambda)=\lambda-\mu$, $\varphi_-(\lambda)=\lambda-\mu-\delta$.
The fusion hierarchy is important to the analysis of the spectra and the derivation of auxiliary functional relations.

The eigenvalues $\Lambda^{[2j]}(\lambda)$ associated to $T^{[2j]}(\lambda)$
also satisfy the functional relations (\ref{fusion}), due to the commutativity
property among transfer matrices. Therefore, we can obtain the eigenvalues at
any fusion level in terms of the first level eigenvalue through the iteration
of the relations (\ref{fusion}) \cite{JSUZUKI99}. 
The eigenvalues of the quantum transfer matrix (\ref{QTMstag}) are given in the form
\bear
\Lambda^{[2j]}(\lambda)=\sum_{m=1}^{2j+1} \lambda_{m}^{(j)}(\lambda), 
\label{eing}
\ear 
\eq
\lambda_{m}^{(j)}(\lambda)=e^{2\beta h(j+1-m)} t_{+,m}^{(j)}(\lambda)
t_{-,m}^{(j)}(\lambda)
\frac{q(\lambda+2\im(\frac{1}{2}+j))q(\lambda_-(\frac{1}{2}+j))}
{q(\lambda+2\im(\frac{3}{2}+j-m))q(\lambda+2\im(\frac{1}{2}+j-m))},
\label{Elambda}
\en
where $\displaystyle t_{\pm,m}^{(j)}(\lambda)=\frac{\varphi_{\pm}(\lambda + 2\im (j\pm\frac{1}{2} -m+1))}{\varphi_{\pm}(\lambda \pm 2\im (j+\frac{1}{2}))}\prod_{l=\pm j + 1 \mp (m-1)}^{j}
\frac{\phi_{\pm}(\lambda \pm 2\im(l-s))}{\phi_{\pm}(\lambda \pm 2\im(l+s))} $ and
$q(\lambda)=\prod_{l=1}^{n}(\lambda-\lambda_{l})$. The corresponding Bethe
ansatz equations can be written as 
\eq
e^{2\beta h}\frac{\phi_{+}(\lambda_l+2\im(s+\frac{1}{2}))
  \phi_{-}(\lambda_l+2\im(s-\frac{1}{2}))}{\phi_{-}(\lambda_l-2\im(s+\frac{1}{2}))
  \phi_{+}(\lambda_l-2\im(s-\frac{1}{2}))}=\prod_{\stackrel{j=1}{j\neq l}}^{n}
\frac{\lambda_l-\lambda_j+2\im}{\lambda_l-\lambda_j-2\im}.
\label{BAeq}
\en
These relations hold for all states. However, for describing the physics of
the model in the thermodynamic limit only the leading eigenstate is
required. The corresponding 
sector has quantum number $n=sN+1$ (remember we are dealing with $N$ copies of
spin $s$ spaces and two spin $1/2$
spaces).  However, one needs to take the Trotter limit $N\rightarrow \infty$
and this cannot be done by a straight forward numerical analysis of the Bethe
ansatz equations. The standard approach allowing for taking the Trotter limit
employs suitable algebraic and analytical properties, in particular auxiliary
functions are used that originate from the analysis of the fusion hierarchy
and from its closure at finite level.  Also analytical properties regarding the
location of zeros and poles are exploited in order to encode the information
of the Bethe ansatz roots in a set of non-linear integral equations. This has
been done for several models and especially for the case of high-spin
chains \cite{JSUZUKI99,KNS013}. Therefore, we present here only the final set
of non-linear integral equations adjusted to the generalized quantum transfer
matrix (\ref{QTMstag}),
\eq
\left(
\begin{array}{c}
\llg{y^{[1]}(\lambda)} \\
\llg{y^{[2]}(\lambda)} \\
\vdots \\
\llg{y^{[2s-1]}(\lambda)} \\
\llg{b(\lambda)} \\
\llg{\bar{b}(\lambda)}
\end{array}\right)=
\left(\begin{array}{c}
{\cal D}(\lambda)  \\
0 \\
\vdots \\
0 \\
 d(\lambda) +\beta \frac{h}{2}  \\
 \\
 d(\lambda) -\beta \frac{h}{2}
\end{array}\right)
+
\widehat{\cal K}*
\left(\begin{array}{c}
\llg{Y^{[1]}(\lambda)} \\
\llg{Y^{[2]}(\lambda)} \\
\vdots \\
\llg{Y^{[2s-1]}(\lambda)} \\
\llg{B(\lambda)} \\
\llg{\bar{B}(\lambda)}
\end{array}\right),
\label{nlie}
\en
where the symbol $*$ denotes the convolution $f*g(\lambda)=\int_{-\infty}^{\infty} f(\lambda-\mu)g(\mu)d\mu$ and 
\bear
{\cal D}(\lambda)&=&\llg\left[\frac{\tanh(\frac{\pi}{4}(\lambda-\mu+\im))}{\tanh(\frac{\pi}{4}(\lambda-\mu-\delta+\im))} \right] \approx -\im \frac{\frac{\pi}{2}}{\cosh(\frac{\pi}{2}(\lambda-\mu))}\cdot \delta, \\
d(\lambda)&=&-\beta J\frac{\frac{\pi}{2}}{2\cosh(\frac{\pi}{2} \lambda )}. 
\ear

The kernel matrix is given explicitly by 
\eq
\widehat{\cal K}(\lambda)= 
\left(\begin{array}{cccccccc}
0 & {\cal K}(\lambda) & 0 & \cdots & 0 & 0 & 0 & 0 \\
{\cal K}(\lambda) & 0 &  {\cal K}(\lambda) &   & \vdots & \vdots & \vdots & \vdots  \\
0 & {\cal K}(\lambda) & 0  &  &  & 0 & 0 & 0 \\
\vdots &  &   &   & 0 & {\cal K}(\lambda) & 0 & 0 \\
0 & 0 & \cdots  & 0 & {\cal K}(\lambda) & 0 & {\cal K}(\lambda) & {\cal K}(\lambda) \\
0 & 0 & \cdots  & 0 & 0 & {\cal K}(\lambda) & {\cal F}(\lambda) & -{\cal F}(\lambda+2\im) \\
0 & 0 & \cdots  & 0 & 0 & {\cal K}(\lambda) & -{\cal F}(\lambda-2\im) & {\cal F}(\lambda) \\
\end{array}\right),
\label{Kernel-x}
\en
where ${\cal K}(\lambda)=\frac{1}{4\cosh( \pi \lambda /2 )}$ and
\bear
{\cal F}(\lambda)=\int_{-\infty}^{\infty}\frac{e^{-|k|-\im k \lambda}}{2 \cosh k} \frac{dk}{2\pi}=\frac{1}{8\pi}\left\{\Psi(-\frac{\im}{4}\lambda) +\Psi(\frac{\im}{4}\lambda) - \Psi(\frac{1}{2}-\frac{\im}{4}\lambda) - \Psi(\frac{1}{2}+\frac{\im}{4}\lambda) \right\}.
\ear
We give the main steps towards the derivation of these equations in Appendix A.

Next we are interested in $\frac{\partial}{\partial\delta}\log\Lambda^{[1]}_0$ for
which we derive an integral expression in terms of functions satisfying a set
of linear integral equations. These functions are
the derivatives of $\log Y^{[k]}, \log B, \log
\bar{B}$ with respect to $\delta$, where $k=2 j=1,2,\cdots,2s-1$. We introduce the functions \cite{KNS013},
\begin{align}
G_{y}^{[k]}(\lambda,\mu)&=\frac{\partial}{\partial\delta}\log
Y^{[k]}(\lambda)\Big|_{\delta=0}, ~
\frac{\partial}{\partial\delta}\log y^{[k]}(\lambda)\Big|_{\delta=0}=[1+{y^{[k]}}^{-1}(\lambda)]\big|_{\delta=0}G_{y}^{[k]}(\lambda,\mu), \nonumber\\
G_{b}(\lambda,\mu)&=\left.\frac{\partial}{\partial\delta}\log B (\lambda)\right|_{\delta=0}, 
~~~  \frac{\partial}{\partial\delta}\log b(\lambda)\Big|_{\delta=0}=[1+b^{-1}(\lambda)]\big|_{\delta=0}G_{b}(\lambda,\mu), \\
G_{\bar{b}}(\lambda,\mu)&=\frac{\partial}{\partial\delta}\log \bar{B}(\lambda)\Big|_{\delta=0}, 
~~~ \frac{\partial}{\partial\delta}\log \bar{b}(\lambda)\Big|_{\delta=0}=[1+\bar{b}^{-1}(\lambda)]\big|_{\delta=0}G_{\bar{b}}(\lambda,\mu).
\nonumber
\end{align}

These satisfy the set of linear integral equations
\eq
\left(
\begin{array}{c}
\left[1+(y^{[1]}(\lambda))^{-1}\right] G_{y}^{[1]}(\lambda,\mu) \\
\left[1+(y^{[2]}(\lambda))^{-1}\right] G_{y}^{[2]}(\lambda,\mu) \\
\vdots \\
\left[1+(y^{[2s-1]}(\lambda))^{-1}\right] G_{y}^{[2s-1]}(\lambda,\mu) \\
\left[1+(b(\lambda))^{-1}\right] G_{b}(\lambda,\mu) \\ 
\left[1+(\bar{b}(\lambda))^{-1}\right] G_{\bar{b}}(\lambda,\mu)
\end{array}\right) 
=
\left(\begin{array}{c}
-\im\frac{\frac{\pi}{2}}{\cosh(\frac{\pi}{2}(\lambda-\mu))} \\ 
0 \\ 
\vdots \\
0 \\
0 \\
0
\end{array}\right) 
 + \widehat{\cal K}*
\left(
\begin{array}{c}
G_{y}^{[1]}(\lambda,\mu) \\ 
G_{y}^{[2]}(\lambda,\mu) \\ 
\vdots \\
G_{y}^{[2s-1]}(\lambda,\mu) \\ 
G_{b}(\lambda,\mu) \\ 
G_{\bar{b}}(\lambda,\mu)
\end{array}\right).
\label{lie}
\en
Next we obtain an explicit expression of the eigenvalue's derivative in terms
of the $G$ functions
\bear
\left.\frac{\partial}{\partial\delta}\log\Lambda_0^{[1]}(\lambda)\right|_{\delta=0}&=&
\int_{-\infty}^{\infty}{d\lambda'\:\frac{1}{4\cosh(\frac{\pi}{2}(\lambda-\lambda'))}}\left\{
G_{y}^{[1]}(\lambda',\mu)+\frac{2\im}{(\lambda'-\mu)^2+1}\right\},\nonumber \\
&=&\int_{-\infty}^{\infty}{d\lambda'\:\frac{1}{4\cosh(\frac{\pi}{2}(\lambda-\lambda'))}} G_{y}^{[1]}(\lambda',\mu)+2\pi \im {\cal F}(\lambda-\mu),\label{derivLam}
\ear
for $|\Im(\lambda)|<1$.

For any high-spin chain \cite{KNS013}, an auxiliary density matrix with two
neighboring spin-$1/2$ quantum spaces (on vertical lines in the statistical
mechanics language) in a sea of $L$ ($\to\infty$) many spin-$s$ objects is
defined. The nearest-neighbour spin-$1/2$ correlators are given by a function
$\omega$, cf.~(\ref{omegaDarstellungDichte1}).  For details see
\cite{AuKl12,KNS013} where a scheme was introduced to derive the two-site
function $\omega$ from a generating function like
\begin{equation}
\omega^{[2s]}(\lambda,\mu)=\frac{1}{2}-\frac{(\lambda-\mu)^2+4}{2\im}\left.\frac{\partial}{\partial\delta}\log\Lambda_0^{[1]}(\lambda)\right|_{\delta=0}.
\end{equation}
With our result (\ref{derivLam}) we have explicitly
\begin{equation}
\omega^{[2s]}(\lambda,\mu)=\frac{1}{2}-\frac{(\lambda-\mu)^2+4}{2\im}\left\{  \int_{-\infty}^{\infty}{d\lambda'\:\frac{G_{y}^{[1]}(\lambda',\mu)}{4\cosh(\frac{\pi}{2}(\lambda-\lambda'))}} +2\pi \im {\cal F}(\lambda-\mu)\right\}.\label{omega}
\end{equation}

\section{Computation of the two-site density matrix}\label{2siteDM}

Due to the $su(2)$ invariance, the two-sites reduced density matrix can be
written as a superposition of the projectors (\ref{su2projector}). Besides
that, thanks to the intertwining symmetry, the density matrix is a symmetric
function $D_2^{[2s]}(\xi_1,\xi_2)=D_2^{[2s]}(\xi_2,\xi_1)$, therefore, one can
write
\eq
D_2^{[2s]}(\xi_1,\xi_2)=\sum_{k=0}^{2 s} \rho_{k}^{[2s]}(\xi_1,\xi_2) \check{P}_k.
\label{D2rep}
\en

Using the above representation for the density matrix (\ref{D2rep}) in the
functional equation qKZ (\ref{qKZ}) specialized to the $2$-site case, implies
the following system of equations,
\eq
\left(\begin{array}{c}
\rho_{0}^{[2s]}(\xi_1-2\im,\xi_2) \\
\rho_{1}^{[2s]}(\xi_1-2\im,\xi_2) \\
\vdots \\
\rho_{2 s}^{[2s]}(\xi_1-2\im,\xi_2)
\end{array}\right)=
A^{[2s]}(\xi) \cdot\left(\begin{array}{c}
\rho_{0}^{[2s]}(\xi_1,\xi_2) \\
\rho_{1}^{[2s]}(\xi_1,\xi_2) \\
\vdots \\
\rho_{2 s}^{[2s]}(\xi_1,\xi_2)
\end{array}\right),
\label{matrixL-qKZ}
\en
where from now on $\xi=\xi_1-\xi_2$ and  $A^{[2s]}(\xi)$ is a $(2s+1)\times
(2s+1)$ matrix which results from the action of the linear operator
$A_2^{[2s]}(\xi_1,\xi_2)$ (\ref{A-qKZ}) on the density matrix
(\ref{D2rep}). We give the explicit form of $A^{[2s]}(\xi)$ for $s$ up to
$3/2$.

\subsection{The spin-$1/2$ case}

The two-site density matrix for the spin-$1/2$ chain is the simplest case as the
density matrix is a linear superposition of projectors onto total spin 0 and
spin 1. The trace condition leaves only one non-trivial parameter to be
determined which is just the $\omega$ function. 

The final result for the physical correlations on two-sites is even simpler
and can be obtained directly from the ground state
energy \cite{HULTHEN}. And this, of course, agrees with the homogeneous limit
of the inhomogeneous density matrix.

So to say for pedagogical reasons, for setting up notations and the way of
reasoning we work out the functional equation satisfied by the two-site
density matrix for the spin-$1/2$ case. The matrix $A^{[1]}(\xi)$ in
(\ref{matrixL-qKZ}) is readily obtained
\eq
A^{[1]}(\xi)=\left(\begin{array}{cc}
-\frac{(\xi-4\im)}{2(\xi+2\im)} & -\frac{3(\xi-4\im)}{2(\xi-2\im)} \\
\frac{\xi}{2(\xi+2\im)} & \frac{\xi}{2(\xi-2\im)} 
\end{array}\right).
\nonumber
\en

The solution for the coefficients $\rho^{[1]}(\xi_1,\xi_2)$ is given by
\bear
\rho_0^{[1]}(\xi_1,\xi_2)&=&\frac{1}{4} -\frac{\omega^{[1]}(\xi_1,\xi_2)}{2}, \\
\rho_1^{[1]}(\xi_1,\xi_2)&=&\frac{1}{4} +\frac{\omega^{[1]}(\xi_1,\xi_2)}{6}.
\ear
where the $\omega^{[1]}(\lambda,\mu)$ (\ref{omega}) is given in terms of the
solution of the integral equations. From (\ref{nlie}) and (\ref{lie}) we read
off the zero-temperature limit $T\rightarrow 0$ of the auxiliary functions,
which allows us to write down the zero-temperature limit of the $\omega$
function
\bear
\lim_{T\rightarrow 0} \omega^{[1]}(\lambda,\mu)&=&\frac{1}{2}-\frac{(\lambda-\mu)^2+4}{8} \Big\{\Psi(-\frac{\im}{4}(\lambda-\mu)) + \Psi(\frac{\im}{4}(\lambda-\mu)) \nonumber \\
&-& \Psi(\frac{1}{2}-\frac{\im}{4}(\lambda-\mu)) -\Psi(\frac{1}{2}+\frac{\im}{4}(\lambda-\mu)) \Big\},
\ear
which is the transcendental part of the correlation
functions \cite{BOOS05}. The computation of the density function for $n>2$ was
successfully done up to $n=8$. These results can be obtained by means of a
suitable ansatz for the coefficients of the transcendental function
$\omega^{[1]}(\lambda,\mu)$ and its products \cite{SST05}.

In the homogeneous limit $\xi_i\rightarrow 0$, we obtain
\bear
\rho_0^{[1]}(0,0)&=&\log2, \\
\rho_1^{[1]}(0,0)&=&\frac{1-\log2}{3}.
\ear

\subsection{The spin-$1$ case}

In the context of general spin-$s$ chains, the first case for which the
physical correlations even for two-sites cannot be obtained just by $su(2)$
symmetry is the spin-$1$ chain \cite{KNS013}. The two-site density matrix is a
superposition of projectors onto total spin 0, 1, and 2. Here we have at hand
the trace condition and a simple correlator (that can be shown to be a linear
combination of $\omega$'s) which are not sufficient to fix the three
coefficients. In this case, the matrix $A^{[2]}(\xi)$ is given by
\eq
A^{[2]}(\xi)=\left(\begin{array}{ccc}
\frac{(\xi-4\im)(\xi-6\im)}{3(\xi+2\im)(\xi+4\im)} & -\frac{(\xi-4\im)(\xi-6\im)}{(\xi-2\im)(\xi+4\im)} & \frac{5(\xi-6\im)}{3(\xi-2\im)} \\
\frac{-\xi(\xi-6\im)}{3(\xi+2\im)(\xi+4\im)} & \frac{\xi(\xi-6\im)}{2(\xi-2\im)(\xi+4\im)} & \frac{5\xi(\xi-6\im)}{6(\xi-2\im)(\xi-4\im)} \\ 
\frac{\xi}{3(\xi+4\im)} & \frac{\xi(\xi+2\im)}{2(\xi-2\im)(\xi+4\im)} & \frac{\xi(\xi+2\im)}{6(\xi-2\im)(\xi-4\im)} 
\end{array}\right).
\nonumber
\en
In \cite{KNS013} solutions to these functional equations were constructed in
the manner of the spin-1/2 factorized form \cite{SST05}. 
According to the fusion principle, the spin-$1$ 2-point function is a special
spin-$1/2$ 4-point function with spectral parameters $\xi_1\pm\im$ and
$\xi_2\pm\im$ in the four local quantum spaces. For finite Trotter number, the
correlations are polynomials in $\xi_1, \xi_2$ divided by
$\Lambda^{[2]}(\xi_1)\Lambda^{[2]}(\xi_2)$. The multi-linear expressions in
terms of $\omega$-functions will be polynomials divided by
$\Lambda^{[1]}(\xi_1+\im)\Lambda^{[1]}(\xi_1-\im)\Lambda^{[1]}(\xi_2+\im)\Lambda^{[1]}(\xi_2-\im)$.
Hence, a scale factor
$N^{[2]}(\xi):=\Lambda^{[2]}(\xi)/\Lambda^{[1]}(\xi+\im)\Lambda^{[1]}(\xi-\im)$
has to be introduced when attempting to satisfy the functional equations and
the analyticity conditions simultaneously. An explicit expression for
$N^{[2]}(\xi)$ in terms of the $\omega$ function is given in the appendix, see
(\ref{N2}).

For satisfying the functional equations (\ref{matrixL-qKZ}), in \cite{KNS013}
an ansatz for $\rho_k^{[2]}(\xi_1,\xi_2)$ in terms of multi-linear expressions
in $\omega$ with rational coefficients in $\xi_1-\xi_2$ was used. The success
of this ansatz depends on the fact that the $\omega$ function itself satisfies
a functional equation closely related to (\ref{matrixL-qKZ}). This equation
was derived by use of the fusion hierarchy \cite{KNS013}. It has the structure
of a $3$-point equation involving the $\omega$ function, which reads, 
\bear
C^{[2]}(\xi_1+\im,\xi_2) +C^{[2]}(\xi_1-\im,\xi_2) =0,
\label{eq74}
\ear
where
\bear
C^{[2]}(\xi_1,\xi_2)&=&\frac{\Omega^{[2]}(\xi_1-\im,\xi_2) + \Omega^{[2]}(\xi_1+\im,\xi_2)
-o(\xi)}{N^{[2]}(\xi_1)},\\
\Omega^{[2]}(\xi_1,\xi_2)&=&2\im\frac{\omega^{[2]}(\xi_1,\xi_2)+\frac{1}{2}}{\xi^2+4}, \\
N^{[2]}(\xi)&=&\frac{3}{4} +\frac{1}{2}\omega^{[2]}(\xi-\im,\xi+\im), \\
o(\xi)&=&\frac{2\im(\xi^2-3)}{(\xi^2+9)(\xi^2+1)}.
\ear
The final result of \cite{KNS013} is
\begin{align}
&\rho_0^{[2]}(\xi_1,\xi_2)=\frac{1}{N^{[2]}(\xi_1)N^{[2]}(\xi_2)} \Big\{ \nonumber  \\
&\frac{1}{16} + \frac{(16 + \xi^2)\omega(\xi_1-\im,\xi_1+\im)}{24\xi^2} - \frac{(4 + \xi^2)\omega(\xi_1-\im,\xi_2+\im)}{6\xi^2} \nonumber \\
 &- \frac{(4 + \xi^2)\omega(\xi_1+\im,\xi_2-\im)}{6\xi^2} - \frac{(4 + \xi^2)^2\omega(\xi_1-\im,\xi_2+\im)\omega(\xi_1+\im,\xi_2-\im)}{36\xi^2} \nonumber\\
 &+ \frac{(16 + \xi^2)\omega(\xi_1-\im,\xi_2-\im)\omega(\xi_1+\im,\xi_2+\im)}{36} + \frac{(16 + \xi^2)\omega(\xi_2-\im,\xi-2+\im)}{24\xi^2} \nonumber \\
 &+ \frac{(16 + \xi^2)\omega(\xi_1-\im,\xi_1+\im)\omega(\xi_2-\im,\xi-2+\im)}{36\xi^2} \Big\}, 
\end{align}
\begin{align}
&\rho_1^{[2]}(\xi_1,\xi_2)=\frac{1}{N^{[2]}(\xi_1)N^{[2]}(\xi_2)} \Big\{  \frac{1}{16} + \frac{(-4 + \xi^2)(16 + \xi^2)\omega(\xi_1-\im,\xi_1+\im)}{24\xi^2(4 + \xi^2)}  \nonumber\\
&- \frac{(16 + \xi^2)\omega(\xi_1-\im,\xi_2-\im)}{12(4 + \xi^2)} + \frac{(2 - \im\xi)\omega(\xi_1-\im,\xi_2+\im)}{3\xi^2} \nonumber \\
&+ \frac{(2 + \im\xi)\omega(\xi_1+\im,\xi_2-\im)}{3\xi^2} + \frac{(4 + \xi^2)(8 + \xi^2)\omega(\xi_1-\im,\xi_2+\im)\omega(\xi_1+\im,\xi_2-\im)}{72\xi^2} \nonumber \\
&- \frac{(16 + \xi^2)\omega(\xi_1+\im,\xi_2+\im)}{12(4 +\xi^2)} - \frac{(8 + \xi^2)(16 + \xi^2)\omega(\xi_1-\im,\xi_2-\im)\omega(\xi_1+\im,\xi_2+\im)}{72(4 + \xi^2)} \nonumber \\
&+ \frac{(-4 + \xi^2)(16 + \xi^2)\omega(\xi_2-\im,\xi-2+\im)}{24\xi^2(4 + \xi^2)} \nonumber \\
&+\frac{(-4 + \xi^2)(16 + \xi^2)\omega(\xi_1-\im,\xi_1+\im)\omega(\xi_2-\im,\xi-2+\im)}{36\xi^2(4 + \xi^2)} \Big\}, 
\end{align}
\begin{align}
&\rho_2^{[2]}(\xi_1,\xi_2)=\frac{1}{N^{[2]}(\xi_1)N^{[2]}(\xi_2)} \Big\{  \frac{1}{16} + \frac{(128 - 20\xi^2 + 5\xi^4)\omega(\xi_1-\im,\xi_1+\im)}{120\xi^2(4 + \xi^2)} \nonumber \\
&+ \frac{(16 + \xi^2)\omega(\xi_1-\im,\xi_2-\im)}{20(4 + \xi^2)} + \frac{(2\im + \xi)(4\im + \xi)\omega(\xi_1-\im,\xi_2+\im)}{30\xi^2} \nonumber \\
&+ \frac{(-2\im + \xi)(-4\im + \xi)\omega(\xi_1+\im,\xi_2-\im)}{30\xi^2} \nonumber\\
&- \frac{(4 + \xi^2)(16 + \xi^2)\omega(\xi_1-\im,\xi_2+\im)\omega(\xi_1+\im,\xi_2-\im)}{360\xi^2} \nonumber \\
&+ \frac{(16 + \xi^2)\omega(\xi_1+\im,\xi_2+\im)}{20(4 + \xi^2)} + \frac{(16 + \xi^2)^2\omega(\xi_1-\im,\xi_2-\im)\omega(\xi_1+\im,\xi_2+\im)}{360(4 + \xi^2)} \nonumber \\
&+ \frac{(128 - 20\xi^2 + 5\xi^4)\omega(\xi_2-\im,\xi-2+\im)}{120\xi^2(4 + \xi^2)} \nonumber \\
&+ \frac{(128 - 20\xi^2 + 5\xi^4)\omega(\xi_1-\im,\xi_1+\im)\omega(\xi_2-\im,\xi-2+\im)}{180\xi^2(4 + \xi^2)} \Big\}.
\end{align}
Taking the zero-temperature limit of the $\omega$-function results in
\bear
\lim_{T\rightarrow 0} \omega^{[2]}(\lambda,\mu)&=&\frac{1}{2}-\frac{(\lambda-\mu)^2+4}{8} \Big\{\Big(\Psi(-\frac{\im}{4}(\lambda-\mu)) + \Psi(\frac{\im}{4}(\lambda-\mu))  \\
&-& \Psi(\frac{1}{2}-\frac{\im}{4}(\lambda-\mu)) -\Psi(\frac{1}{2}+\frac{\im}{4}(\lambda-\mu)) \Big) -\frac{\pi(\lambda-\mu)}{2\sinh(\frac{\pi}{2}(\lambda-\mu))} \Big\}. \nonumber
\ear
Taking furthermore the homogeneous limit and using the zero-temperature
$\omega$-function, the coefficients of the projection operators are obtained
\bear
\rho_0^{[2]}(0,0)&=&-\frac{8}{9}+\frac{4}{27}\pi^2, \nonumber\\
\rho_1^{[2]}(0,0)&=&\frac{14}{9}-\frac{4}{27}\pi^2, \\
\rho_2^{[2]}(0,0)&=&-\frac{5}{9}+\frac{8}{135}\pi^2. \nonumber
\ear
In this case, only $\pi^2$ appears which is related to the zeta function
value $\zeta(2)=\frac{\pi^2}{6}$. For $s=1$, it was also possible to deal with
the three site density matrix \cite{KNS013}, whose results are given in terms of
$\pi^2,\pi^4$ and $\pi^6$ or alternatively $\zeta(2),\zeta(4)$ and $\zeta(6)$.

\subsection{The spin-$3/2$ case}

The case of $s=3/2$ is the subject of our new work. This case is non-trivial as
it is computationally much more involved than the lower spin cases. This is
because the ansatz for $\rho_k^{[3]}(\xi_1,\xi_2)$ is built upon the fusion of
spin-$1/2$ objects such that the spin-$s$ case with $n$ sites is as involved
as the spin-$1/2$ case with $2 s n$ sites. Here, the matrix $A^{[3]}(\xi)$ of
the functional equation (\ref{matrixL-qKZ}) of the two-site density operator
reads 
\eq 
A^{[3]}(\xi)=\left(\begin{array}{cccc}
  \frac{-(\xi-4\im)(\xi-6\im)(\xi-8\im)}{4(\xi+2\im)(\xi+4\im)(\xi+6\im)} &
  \frac{3(\xi-4\im)(\xi-6\im)(\xi-8\im)}{4(\xi-2\im)(\xi+4\im)(\xi+6\im)} &
  \frac{-5(\xi-6\im)(\xi-8\im)}{4(\xi-2\im)(\xi+6\im)} &
  \frac{7(\xi-8\im)}{4(\xi-2\im)}
  \\ \frac{\xi(\xi-6\im)(\xi-8\im)}{4(\xi+2\im)(\xi+4\im)(\xi+6\im)} &
  \frac{-11\xi(\xi-6\im)(\xi-8\im)}{20(\xi-2\im)(\xi+4\im)(\xi+6\im)} &
  \frac{\xi(\xi-6\im)(\xi-8\im)}{4(\xi-2\im)(\xi-4\im)(\xi+6\im)} &
  \frac{21\xi(\xi-8\im)}{20(\xi-2\im)(\xi-4\im)}
  \\ \frac{-\xi(\xi-8\im)}{4(\xi+4\im)(\xi+6\im)} &
  \frac{3\xi(\xi+2\im)(\xi-8\im)}{20(\xi-2\im)(\xi+4\im)(\xi+6\im)} &
  \frac{3\xi(\xi+2\im)(\xi-8\im)}{4(\xi-2\im)(\xi-4\im)(\xi+6\im)} &
  \frac{7\xi(\xi+2\im)(\xi-8\im)}{20(\xi-2\im)(\xi-4\im)(\xi-6\im)}
  \\ \frac{\xi}{4(\xi+6\im)} & \frac{9\xi(\xi+2\im)}{20(\xi-2\im)(\xi+6\im)} &
  \frac{\xi(\xi+2\im)(\xi+4\im)}{4(\xi-2\im)(\xi-4\im)(\xi+6\im)} &
  \frac{\xi(\xi+2\im)(\xi+4\im)}{20(\xi-2\im)(\xi-4\im)(\xi-6\im)}
\end{array}\right).
\label{L3}
\en
We use an ansatz for the solution in terms of multilinear expressions in
$\omega$ functions with rational coefficients. 
The expressions contain up to trilinear combinations of $\omega$-functions
with arguments $\xi_1$, $\xi_1\pm 2\im$, $\xi_2$, $\xi_2\pm 2\im$.
We use a scale factor
$N^{[3]}(\xi):=\Lambda^{[3]}(\xi)/\Lambda^{[1]}(\xi+2\im)\Lambda^{[1]}(\xi)\Lambda^{[1]}(\xi-2\im)$
that can be expressed just in terms of $\omega$ functions, see (\ref{Bfusion4}).
The analogue of the functional equation (\ref{eq74}) is given by,
\bear
C^{[3]}(\xi_1+\im,\xi_2) +C^{[3]}(\xi_1-\im,\xi_2) =\frac{1}{2}\left[\left(\frac{1}{\xi+\im} -\frac{1}{\xi-\im}\right) +\left(\frac{1}{\xi+5\im} -\frac{1}{\xi-5\im}\right)\right],
\label{analogueof74}
\ear
where 
\begin{align}
C^{[3]}(\xi_1,\xi_2)&=\Big\{-\left(\frac{3}{2}+\omega^{[3]}(\xi_1,\xi_1+2\im)\right)\Omega^{[3]}(\xi_1-2\im,\xi_2)\nonumber  \\
&-2 \Omega^{[3]}(\xi_1,\xi_2)-\left(\frac{3}{2}+\omega^{[3]}(\xi_1-2\im,\xi_1)\right)\Omega^{[3]}(\xi_1+2\im,\xi_2) \nonumber \\
&+\frac{1}{2}\left[-\frac{1}{\xi} +\frac{1}{\xi-2\im} -\frac{1}{\xi-4\im} +\frac{1}{\xi+2\im} \right]\omega^{[3]}(\xi_1-2\im,\xi_1) \nonumber \\
&-\frac{1}{2}\left[-\frac{1}{\xi} +\frac{1}{\xi-2\im} -\frac{1}{\xi+4\im} +\frac{1}{\xi+2\im} \right]\omega^{[3]}(\xi_1,\xi_1+2\im) \\
&+\frac{2\im}{\xi^2+16}\Big\}/N^{[3]}(\xi_1), \nonumber\\
\Omega^{[3]}(\xi_1,\xi_2)&=2\im\frac{\omega^{[3]}(\xi_1,\xi_2)+\frac{1}{2}}{\xi^2+4}, \\
 N^{[3]}(\xi)&=1+\omega^{[3]}(\xi,\xi+2\im) + \omega^{[3]}(\xi-2\im,\xi).
\end{align}
This equation is obtained from the fusion hierarchy and holds true at certain
values where $\phi(\xi_1)=0$. 
Another equation that is essential for
consistently satisfying the functional equations is (\ref{consist2}) which has
no counterpart in the spin-$1$ case.  We give some details of the derivations
in Appendix B.

By use of (\ref{matrixL-qKZ}), (\ref{L3}) and (\ref{analogueof74}), we obtain
for $\widetilde{\rho}_i:=N^{[3]}(\xi_1) N^{[3]}(\xi_2)\rho_i^{[3]}(\xi_1,\xi_2) $,
\begin{footnotesize}
\begin{align}
&\widetilde{\rho}_0(\xi_1,\xi_2)=\frac{1}{16} ( 1+\omega(\xi_1^-,\xi_1)+\omega(\xi_1,\xi_1^+) )  ( 1+\omega(\xi_2^-,\xi_2)+\omega(\xi_2,\xi_2^+) ) -\frac{1}{4}{\frac{({\xi}^{2}+16)\omega(\xi_1,\xi_2)}{(\xi^2+4)}}\nonumber\\
 &+\frac{3}{2}\left[{\frac{\omega(\xi_1,\xi_1^+)}{\xi(\xi+2\im)}}+{\frac{\omega(\xi_1^-,\xi_1)}{\xi(\xi-2\im)}} +{\frac{\omega(\xi_2,\xi_2^+)}{\xi(\xi-2\im)}}+{\frac {\omega(\xi_2^-,\xi_2)}{\xi(\xi+2\im)}}\right]\nonumber\\
 &-\frac{3}{16}\left[{\frac { ( \xi+2\im)  ( \xi-4\im ) \omega(\xi_1^-,\xi_2^+)}{\xi ( \xi-2\im ) }} +{\frac{( \xi-2\im )  (\xi+ 4\im ) \omega(\xi_1^+,\xi_2^-)}{\xi ( \xi+2\im ) }}\right]\nonumber\\
 &-\frac{3}{5}\left[{\frac{(-3 \xi^{3}+2\im{\xi}^{2}-8 \xi-48\im) \omega(\xi_1,\xi_1^+)\omega(\xi_2^-,\xi_2)}{{\xi}^{3}(\xi^2+4)}}+{\frac{(-3 \xi^{3}-2\im{\xi}^{2}-8 \xi+48\im) \omega(\xi_1^-,\xi_1)\omega(\xi_2,\xi_2^+)}{{\xi}^{3}(\xi^2+4)}}\right]\nonumber\\
 &+\frac{9}{5}{\frac { (\xi^2-4)}{  {\xi}^{2} ( \xi^2+4)   }}\left[\omega(\xi_1^-,\xi_1)\omega(\xi_2^-,\xi_2) + \omega(\xi_1,\xi_1^+)\omega(\xi_2,\xi_2^+)\right]\nonumber\\
 &-{\frac{3}{80}}\,{\frac{(\xi^2+16)(\xi-6\im)}{{\xi}^{2}(\xi+2\im)}}\left[\omega(\xi_1^+,\xi_2^-)\omega(\xi_2,\xi_2^+) + \omega(\xi_1^-,\xi_1)\omega(\xi_1^+,\xi_2^-)\right]\nonumber
 \end{align}
 \begin{align}   
 &-{\frac{3}{80}}\,{\frac{(\xi^2+16)(\xi+6\im)}{{\xi}^{2}(\xi-2\im)}} \left[\omega(\xi_1^-,\xi_2^+)\omega(\xi_2^-,\xi_2) + \omega(\xi_1,\xi_1^+)\omega(\xi_1^-,\xi_2^+)\right]\nonumber\\
 &-{\frac{3}{80}}\,{\frac{({\xi}^{2}+36)({\xi}^{2}+16)}{{\xi}^{2}(\xi^2+4)}} \left[\omega(\xi_1^+,\xi_2^+)\left(\omega(\xi_1^-,\xi_1)+ \omega(\xi_2^-,\xi_2)\right)+\omega(\xi_1^-,\xi_2^-)\left(\omega(\xi_1,\xi_1^+) + \omega(\xi_2,\xi_2^+)\right) \right]\nonumber\\
 &+{\frac{1}{1280}}\,{({\xi}^{2}+16)^{2}\omega(\xi_1^-,\xi_2^+)\omega(\xi_1^+,\xi_2^-)}\nonumber\\
 &-{\frac{1}{1280}}\,{\frac{(\xi^2+36)(\xi-2\im)^{2}(\xi+4\im)^{2}}{{\xi}^{2}}}\left[\omega(\xi_1,\xi_2^+)\omega(\xi_1^+,\xi_2^-)  +\omega(\xi_1^-,\xi_2)\omega(\xi_1^+,\xi_2^-)\right]\nonumber\\
 &-{\frac{1}{1280}}\,{\frac{(\xi^2+36)(\xi+2\im)^{2}(\xi-4\im)^{2}}{{\xi}^{2}}}\left[\omega(\xi_1,\xi_2^-)\omega(\xi_1^-,\xi_2^+)+\omega(\xi_1^+,\xi_2)\omega(\xi_1^-,\xi_2^+)\right]\nonumber\\
 &-{\frac{1}{1280}}\,{\frac{({\xi}^{3}-6\im{\xi}^{2}-24 \xi-96\im)(\xi^2+4)(\xi-6\im)(\xi+4\im)\omega(\xi_1,\xi_2^+)\omega(\xi_1^-,\xi_2)}{ {\xi}^{3}}}\nonumber\\
 &-{\frac{1}{1280}}\,{\frac{({\xi}^{3}+6\im{\xi}^{2}-24 \xi+96\im)(\xi^2+4)(\xi+6\im)(\xi-4\im)\omega(\xi_1,\xi_2^-)\omega(\xi_1^+,\xi_2)}{ {\xi}^{3}}}\nonumber\\
 &+{\frac{1}{1280}}\,{\frac{({\xi}^{3}-6\im{\xi}^{2}-24 \xi-96\im)(\xi^2+16)(\xi-4\im)\omega(\xi_1,\xi_2)\omega(\xi_1^-,\xi_2^+)}{\xi(\xi-2\im)}}\nonumber\\
 &+{\frac{1}{1280}}\,{\frac{({\xi}^{3}+6\im{\xi}^{2}-24\,\xi+96\im)(\xi^2+16)(\xi+4\im)\omega(\xi_1,\xi_2)\omega(\xi_1^+,\xi_2^-)}{\xi(\xi+2\im) }}\nonumber\\
 &+{\frac{1}{1280}}\,{\frac{({\xi}^{2}+36)({\xi}^{2}+16)(\xi-2\im)(\xi+6\im)}{{\xi}^{2}}}\left[\omega(\xi_1^+,\xi_2)\omega(\xi_1^-,\xi_2^-) + \omega(\xi_1^+,\xi_2^+)\omega(\xi_1,\xi_2^-)\right]\nonumber\\
 &+{\frac{1}{1280}}\,{\frac{({\xi}^{2}+36)({\xi}^{2}+16)(\xi+2\im)(\xi-6\im)}{{\xi}^{2}}}\left[\omega(\xi_1^-,\xi_2)\omega(\xi_1^+,\xi_2^+)+ \omega(\xi_1^-,\xi_2^-)\omega(\xi_1,\xi_2^+)\right]\nonumber\\
 &+{\frac{1}{1280}}\,{\frac{({\xi}^{2}+36)({\xi}^{2}+16)({\xi}^{2}+4)}{{\xi}^{2}}}\left[\omega(\xi_1,\xi_2^-)\omega(\xi_1^-,\xi_2)+\omega(\xi_1,\xi_2^+)\omega(\xi_1^+,\xi_2)\right]\nonumber\\
 &-{\frac{1}{1280}}\,{\frac{({\xi}^{2}+36)({\xi}^{2}+16)^{2}}{(\xi^2+4)}} \left[\omega(\xi_1^-,\xi_2^-)\omega(\xi_1^+,\xi_2^+) +\omega(\xi_1,\xi_2)\omega(\xi_1^+,\xi_2^+)+ \omega(\xi_1^-,\xi_2^-)\omega(\xi_1,\xi_2)\right] \nonumber\\
 &-\frac{1}{20}{\frac{(\xi^2+36)(\xi-4\im)}{{\xi}^{3}}}\left[\omega(\xi_1^+,\xi_2)\omega(\xi_1^-,\xi_1) +\omega(\xi_1,\xi_2^-)\omega(\xi_2,\xi_2^+)\right]\nonumber\\
 &-\frac{1}{20}{\frac{(\xi^2+36)(\xi+4\im)}{{\xi}^{3}}}\left[\omega(\xi_1^-,\xi_2)\omega(\xi_1,\xi_1^+)+\omega(\xi_1,\xi_2^+)\omega(\xi_2^-,\xi_2)\right]\nonumber\\
 &-\frac{1}{40}\left[{\frac{(\xi^2+16)(\xi-6\im)\omega(\xi_1^+,\xi_2^-)\omega(\xi_1^-,\xi_1)\omega(\xi_2,\xi_2^+)}{{\xi}^{2}(\xi+2\im)}}+{\frac{(\xi^2+16)(\xi+6\im)\omega(\xi_1^-,\xi_2^+)\omega(\xi_1,\xi_1^+)\omega(\xi_2^-,\xi_2)}{{\xi}^{2}(\xi-2\im)}}\right]\nonumber\\
 &-\frac{1}{40}{\frac{({\xi}^{2}+36)({\xi}^{2}+16)}{{\xi}^{2}(\xi^2+4)}}\left[\omega(\xi_1^+,\xi_2^+)\omega(\xi_1^-,\xi_1)\omega(\xi_2^-,\xi_2)+\omega(\xi_1^-,\xi_2^-)\omega(\xi_1,\xi_1^+)\omega(\xi_2,\xi_2^+)\right]\nonumber\\
 &-{\frac{1}{3840}}\,{\frac{({\xi}^{2}+6)(\xi^2+4)}{{\xi}^{2}}} \Big[(\xi-6\im)(\xi+4\im)^{2}(\xi-2\im)\omega(\xi_1^+,\xi_2^-)\omega(\xi_1,\xi_2^+)\omega(\xi_1^-,\xi_2) \nonumber\\
 &+ (\xi+6\im)(\xi-4\im)^{2}(\xi+2\im)\omega(\xi_1^-,\xi_2^+)\omega(\xi_1,\xi_2^-)\omega(\xi_1^+,\xi_2)\Big]\nonumber\\
 &+{\frac{1}{3840}}\,{\frac{({\xi}^{2}+36)({\xi}^{2}+16)({\xi}^{2}+6)({\xi}^{2}+4)}{{\xi}^{2}}}\left[\omega(\xi_1^+,\xi_2)\omega(\xi_1,\xi_2^+)\omega(\xi_1^-,\xi_2^-)+\omega(\xi_1^-,\xi_2)\omega(\xi_1,\xi_2^-)\omega(\xi_1^+,\xi_2^+)\right]\nonumber\\
 &-{\frac{1}{3840}}\,{\frac{({\xi}^{2}+36)({\xi}^{2}+16)^{2}({\xi}^{2}+6)\omega(\xi_1^-,\xi_2^-)\omega(\xi_1,\xi_2)\omega(\xi_1^+,\xi_2^+)}{(\xi^2+4)}}\nonumber\\
 &+{\frac{1}{3840}}\,{({\xi}^{2}+16)^{2}({\xi}^{2}+6)\omega(\xi_1^-,\xi_2^+) \omega(\xi_1,\xi_2)\omega(\xi_1^+,\xi_2^-)},
\end{align}
\end{footnotesize}
where for convenience we abbreviated
$\omega(\xi_1,\xi_2):=\omega^{[3]}(\xi_1,\xi_2)$ and
$\xi_i^{\pm}=\xi_i \pm 2\im$. The remaining coefficients
are listed in Appendix C.

Based on the knowledge of the coefficients $\rho_k^{[3]}(\xi_1,\xi_2)$, we can easily calculate any nearest neighbour correlators after the homogeneous limit is taken. For instance, in the case of the internal energy $e=\langle {\cal H}^{[3]} \rangle=-T^2\frac{\partial}{\partial T} (f/T)$ at zero magnetic field, we have
\eq
e=J\left(\frac{11}{6}\rho^{[3]}_0(0,0) + \frac{5}{2}\rho_1^{[3]}(0,0) +\frac{5}{3}\rho_2^{[3]}(0,0)\right).
\label{ienergy}
\en

At zero temperature, one has the transcendental function given by 
\bear
\lim_{T\rightarrow 0} \omega^{[3]}(\lambda,\mu)&=&\frac{1}{2}-\frac{(\lambda-\mu)^2+4}{8} \Big\{\Big(\Psi(-\frac{\im}{4}(\lambda-\mu)) + \Psi(\frac{\im}{4}(\lambda-\mu))   \nonumber\\
&-& \Psi(\frac{1}{2}-\frac{\im}{4}(\lambda-\mu)) -\Psi(\frac{1}{2}+\frac{\im}{4}(\lambda-\mu)) \Big) \\
&-&\frac{2\pi}{\sinh(\frac{2\pi}{5})}\frac{\sinh(\frac{\pi}{10}(\lambda-\mu))}{\sinh(\frac{\pi}{2}(\lambda-\mu))} \Big\}, \nonumber
\ear
which in the homogeneous limit produces the following results,
\bear
\rho_0^{[3]}(0,0)&=&\left(3-\frac{3}{10}\pi^2 \right) +\left(-\frac{21}{4} +\frac{3}{5}\pi^2 \right)\log2,  \nonumber\\
\rho_1^{[3]}(0,0)&=&\left(-\frac{183}{20}+\frac{47}{50}\pi^2 \right) +\left(\frac{291}{20}-\frac{37}{25}\pi^2 \right)\log2, \nonumber\\
\rho_2^{[3]}(0,0)&=&\left(\frac{429}{40}-\frac{27}{25}\pi^2 \right) +\left(-\frac{309}{20}+\frac{39}{25}\pi^2 \right)\log2, \label{rho2}\\
\rho_3^{[3]}(0,0)&=&\left(-\frac{161}{40}+\frac{72}{175}\pi^2 \right) +\left(\frac{111}{20}-\frac{99}{175}\pi^2 \right)\log2. \nonumber
\ear
It is worth to note that for $s=3/2$ and two-sites $\pi^2$ and $\log 2$
appear. This has a structure which is different from the above results for
$s=1/2,1$. For the cases $s=1/2$ and $s=1$ all known data indicate that the
results are given by combinations of $\log 2$, $\zeta(3),\zeta(5),\cdots$ or
in powers of $\pi^2$ (zeta function values of even integer arguments). In the
present half-odd integer spin case, we do not find expressions just in terms
of zeta function values of odd integer arguments, we find a mixture, and this
already for the two-site correlation functions.

At last, in order to compute the energy at zero temperature  we insert
(\ref{rho2}) into (\ref{ienergy}). The result is simply given as
\eq
e_{T=0}=J\left(\frac{1}{2} + \log 2 \right),
\en
which is in agreement with the literature \cite{TAK}.

We can also compute spin correlators, e.g at zero temperature we find
\bear
\langle S^z_i S^z_{i+1} \rangle=\left(-\frac{105}{8}+\frac{13}{10}\pi^2 \right) +\left(15 -\frac{8}{5}\pi^2 \right)\log2 \approx -0.843048...
\ear
Due to isotropy, one has $\langle S^x_i S^x_{i+1} \rangle=\langle S^y_i
S^y_{i+1} \rangle=\langle S^z_i S^z_{i+1} \rangle$ implying $\langle
\vec{S}_i\cdot \vec{S}_{i+1} \rangle =3 \langle S^z_i S^z_{i+1} \rangle
\approx -2.529144...$.

\section{Conclusion}
\label{CONCLUSION}

We have exploited the new approach developed in \cite{KNS013} to obtain
further results for correlation functions of high-spin $su(2)$ chains. It is
based on the discrete functional relation of quantum Knizhnik-Zamolodchikov
type and the fusion procedure.

We obtained the general two-site correlation function for the integrable
spin-$3/2$ chain. Surprisingly the result is given in terms of $\log2$ and
$\pi^2$, which can be seen as zeta function values of even and odd arguments.
This structure is very different from that of the famous results for the
spin-$1/2$ case \cite{BOKO03,BST05,SST05} and from that of the recently studied
spin-$1$ chain \cite{KNS013}. For the spin-$1/2$ case the result is given
only in terms of zeta function values with odd arguments and for the spin-$1$
case only zeta function values of even arguments appear.

Although having shown that this approach is viable for an explicit computation
of the correlation functions, the direct application to more spins becomes
quickly cumbersome. Finding an alternative, elegant computational tool would
be highly desirable. Also, the generalization of our approach to higher rank
spin chains would be very interesting. We hope to come back to these
scientific issues in the near future.

\section*{Acknowledgments}

The authors thank the S\~ao Paulo Research Foundation (FAPESP) for financial support through the grants 2015/07780-7 and 2015/01643-8.
A.K. acknowledges the hospitality of Universidade Federal de S\~ao Carlos where the
main part of the work has been carried out. He also acknowledges support by
CNRS and the hospitality of Universit\'e Pierre et Marie Curie. 
G.A.P. Ribeiro thanks the Simons Center for Geometry and Physics of Stony Brook University
and the organizers of the scientific program "Statistical mechanics and combinatorics" 
for hospitality and support during part of this work.

\section*{\bf Appendix A: Algebraic relations and auxiliary functions}
\setcounter{equation}{0}
\renewcommand{\theequation}{A.\arabic{equation}}

For our purposes it is convenient to deal with polynomials rather than with
rational functions. For this reason, we slightly change the normalization of
the $R$-matrix, which naturally results in transfer matrix eigenvalues with
different normalization. The relation among the eigenvalues with different
normalization is given by,
\eq
\Lambda_{2j}(\lambda)=\prod_{l=1}^{2j} \prod_{\sigma=\pm}\phi_{\sigma}(\lambda +\im 2\sigma  (s-j+l)) \varphi_{\sigma}(\lambda +\im 2 \sigma (\frac{1}{2}-j+ l)) \Lambda^{[2j]}(\lambda),
\label{diff-norm}
\en
Using this new normalization, we have the following modified expressions:

Fusion hierarchy
\begin{align}
&T_{2j}(\lambda)T_{1}(\lambda+\im(2j+1))=
T_{2j+1}(\lambda+\im ) +
\hat{\chi}(\lambda+\im (2 j))T_{2j-1}(\lambda-\im), \nonumber\\
&T_{0}(\lambda)= \mbox{Id}, ~~ j=1/2,1,3/2\dots 
\label{fusion2}
\end{align}
where 
\begin{align}
\hat{\chi}(\lambda)&=\phi(\lambda-\im(2 s+1))\phi(\lambda+\im(2 s+1)) \varphi(\lambda-2\im)\varphi(\lambda+2\im),\nonumber\\
\phi(\lambda)&=\phi_{+}(\lambda + \im)\phi_{-}(\lambda - \im),\nonumber\\
\varphi(\lambda)&=\varphi_{+}(\lambda + \im)\varphi_{-}(\lambda - \im).
\end{align}
The $T$-system
\begin{align}
&T_{2j}(\lambda+\im)T_{2j}(\lambda-\im)=
T_{2j-1}(\lambda ) T_{2j+1}(\lambda ) +
f_j(\lambda)\mbox{Id}, ~ j=1/2,1,3/2\dots 
\label{Tsystem2}
\end{align}
where $f_j(\lambda)=\prod_{l=-j+1}^{j} \prod_{\sigma=\pm} \phi(\lambda + \im \sigma (2s+2l) ) \varphi(\lambda + \im \sigma (1+2l) )$.

Eigenvalue expressions
\bear
\Lambda_{2j}(\lambda)=\sum_{m=1}^{2j+1} \hat{\lambda}_{m}^{(j)}(\lambda), 
\label{eing2}
\ear 
\eq
\hat{\lambda}_{m}^{(j)}(\lambda)=e^{2\beta h(j+1-m)} \hat{t}_{+,m}^{(j)}(\lambda)
\hat{t}_{-,m}^{(j)}(\lambda)
\frac{q(\lambda+2\im(\frac{1}{2}+j))q(\lambda-2\im(\frac{1}{2}+j))}
{q(\lambda+2\im(\frac{3}{2}+j-m))q(\lambda+2\im(\frac{1}{2}+j-m))},
\label{Elambda2}
\en
where $\displaystyle \hat{t}_{\pm,m}^{(j)}(\lambda)=\prod_{l=\pm j + 1 \mp (m-1)}^{j}
\phi(\lambda \pm \im(2l-2s-1)) \varphi(\lambda \pm \im(2l-2)) $ and
$q(\lambda)=\prod_{l=1}^{n}(\lambda-\lambda_{l})$. The corresponding Bethe ansatz equations can
be written as 
\eq
e^{2\beta h}\frac{\phi(\lambda_l+\im(2s))}{\phi(\lambda_l-\im(2s))}=\prod_{\stackrel{j=1}{j\neq l}}^{n}
\frac{\lambda_l-\lambda_j+2\im}{\lambda_l-\lambda_j-2\im}.
\label{BAeq2}
\en

This allows us to define a suitable set of auxiliary functions as
\eq
y^{[2j]}(\lambda) =\frac{\Lambda_{2j-1}(\lambda)\Lambda_{2j+1}(\lambda)}{f_j(\lambda)}, ~~ j=\frac{1}{2},\dots,s-\frac{1}{2}.
\label{auxy2}
\en
and
\bear 
b(\lambda)&=&\frac{\hat{\lambda}_2^{(s)}(\lambda+\im) + \dots + \hat{\lambda}_{2s+1}^{(s)}(\lambda+\im)}{\hat{\lambda}_{1}^{(s)}(\lambda+\im)},\label{auxb}\\ 
\bar{b}(\lambda)&=&\frac{\hat{\lambda}_1^{(s)}(\lambda-\im) + \dots + \hat{\lambda}_{2s}^{(s)}(\lambda-\im)}{\hat{\lambda}_{2s+1}^{(s)}(\lambda-\im)}. \label{auxbbar}
\ear 
In addition to this, we define $B(\lambda):=1+b(\lambda)$, $\bar{B}(\lambda):=1+\bar{b}(\lambda)$ and $Y^{[2j]}(\lambda):=1+y^{[2j]}(\lambda)$.

According the previous definition, we note that $B(\lambda)=\frac{\Lambda_{2s}(\lambda+\im)}{\hat{\lambda}_{1}^{(s)}(\lambda+\im)}$
and $\bar{B}(\lambda)=\frac{\Lambda_{2s}(\lambda-\im)}{\hat{\lambda}_{2s+1}^{(s)} (\lambda-\im)}$ with product $B(\lambda)\bar{B}(\lambda) =
Y^{[2s]}(\lambda)$. This implies for the first $(2 s-1)$ functional relations
\begin{align}
&y^{[2j]}(\lambda+\im)y^{[2j]}(\lambda-\im)=Y^{[2j-1]}(\lambda)Y^{[2j+1]}(\lambda) ~ \mbox{for } j=\frac{1}{2},1,\dots,s-1, \label{yrelat}\\
&y^{[2s-1]}(\lambda+\im)y^{[2s-1]}(\lambda-\im)=Y^{[2s-2]}(\lambda)B(\lambda)\bar{B}(\lambda).  
\label{yBrelat}
\end{align}

Therefore, we end up in the following set of algebraic relations,
\begin{align}
&b(\lambda)=\frac{q(\lambda+\im(2s+2))}{q(\lambda-\im 2s)}\frac{e^{-\beta h(2s+1)}\phi(\lambda)\varphi(\lambda+\im(2s-1))\Lambda_{2s-1}(\lambda)}{\prod_{l=1}^{2 s} \phi(\lambda+\im 2l ) \varphi(\lambda+\im (2s-2l+3) )},
\label{baux} \\
&\bar{b}(\lambda)=\frac{q(\lambda-\im(2s+2))}{q(\lambda+\im 2s)}\frac{e^{\beta h(2s+1)}\phi(\lambda)\varphi(\lambda-\im(2s-1))\Lambda_{2s-1}(\lambda)}{\prod_{l=1}^{2 s} \phi(\lambda-\im 2l ) \varphi(\lambda-\im (2s-2l+3) )},
\label{bbaraux}
\end{align}
In this way, it is evident that $b(\lambda)$, $\bar{b}(\lambda)$ are related to
$\Lambda_{2s-1}(\lambda)$.

Moreover, $\Lambda_{2s-1}(\lambda)$ is related to $Y^{[2s-1]}(\lambda)$
through the definition of the $y$-functions. This relation can be written as
\eq
\Lambda_{2s-1}(\lambda+\im)\Lambda_{2s-1} (\lambda-\im)=f_{s-\frac{1}{2}}(\lambda)Y^{[2s-1]}(\lambda). 
\label{auxY}
\en

Now we have all the ingredients to derive the non-linear integral
equations \cite{JSUZUKI99,RK2008,KNS013}. The main idea is to compute the
Fourier transform of the logarithm of the above defined relations
(\ref{yrelat}-\ref{auxY}). This allows us to get rid of the Bethe ansatz roots
by eliminating the function $q(\lambda)$. After a long but straightforward calculation,
where we take the inverse Fourier transform of the transformed auxiliary
functions, we finally obtain the non-linear integral equations (\ref{nlie}).

\section*{\bf Appendix B: Functional equations for the basic functions for $s=3/2$}
\setcounter{equation}{0}
\renewcommand{\theequation}{B.\arabic{equation}}

From the fusion hierarchy (\ref{fusion2}) we have the following explicit relations,
\begin{align}
&\Lambda_{2}(\lambda )=\Lambda_{1}(\lambda-\im)\Lambda_{1}(\lambda+ \im)- \phi(\lambda-4\im)\phi(\lambda+4\im) \varphi(\lambda-2\im)\varphi(\lambda+2\im),\\
&\Lambda_{3}(\lambda )=\Lambda_{2}(\lambda-\im)\Lambda_{1}(\lambda+2\im) - \phi(\lambda-3\im)\phi(\lambda+5\im) \varphi(\lambda-\im)\varphi(\lambda+3\im)\Lambda_{1}(\lambda-2\im), 
\end{align}
Besides that, from the eigenvalue expression (\ref{eing2}) we see that
\begin{align}\Lambda_{3}(\lambda-\im)\Lambda_{3}(\lambda+ \im)&=G(\lambda)  \left[\varphi(\lambda)\varphi(\lambda-2\im)\varphi(\lambda+2\im)\varphi(\lambda+4\im)\varphi(\lambda)\varphi(\lambda-4\im) \right]\nonumber \\
&+ \phi(\lambda)\widetilde{\Lambda}(\lambda),\label{inversion}
\end{align}
where $G(\lambda)=\prod_{l=1}^3  \prod_{\sigma=\pm}\phi(\lambda +\sigma \im 2 l)$ is independent of $\delta$.

We divide all of the above equations by the $\varphi$-function with the
respective arguments on the right
\begin{align}
&\frac{\Lambda_{2}(\lambda )}{\varphi(\lambda-2\im)\varphi(\lambda+2\im)}=\frac{\Lambda_{1}(\lambda-\im)\Lambda_{1}(\lambda+ \im)}{\varphi(\lambda-2\im)\varphi(\lambda+2\im)}- \phi(\lambda-4\im)\phi(\lambda+4\im),\label{interm1}\\
&\frac{\Lambda_{3}(\lambda )}{\varphi(\lambda-\im)\varphi(\lambda+3\im)}=\frac{\Lambda_{2}(\lambda-\im)\Lambda_{1}(\lambda+2\im)}{\varphi(\lambda-\im)\varphi(\lambda+3\im)} - \phi(\lambda-3\im)\phi(\lambda+5\im) \Lambda_{1}(\lambda-2\im), \label{interm2}
\end{align}
\begin{align}
&\frac{\Lambda_{3}(\lambda-\im)}{\varphi(\lambda-2\im)\varphi(\lambda+2\im)} \cdot\frac{\Lambda_{3}(\lambda+ \im)}{\varphi(\lambda)\varphi(\lambda+4\im)} \cdot \frac{1}{\varphi(\lambda)\varphi(\lambda-4\im) }= \nonumber \\
&=G(\lambda)+ \frac{\phi(\lambda)\widetilde{\Lambda}(\lambda)}{\varphi(\lambda)\varphi(\lambda-2\im)\varphi(\lambda+2\im)\varphi(\lambda+4\im)\varphi(\lambda)\varphi(\lambda-4\im)}.
\label{interm}
\end{align}

From the last equation we derive
\begin{align}
&\frac{\partial}{\partial\delta}\log\frac{\Lambda_{3}(\lambda-\im)}{\varphi(\lambda-2\im)\varphi(\lambda+2\im)} +\frac{\partial}{\partial\delta}\log \frac{\Lambda_{3}(\lambda+ \im)}{\varphi(\lambda)\varphi(\lambda+4\im)} +\frac{\partial}{\partial\delta}\log \frac{1}{\varphi(\lambda)\varphi(\lambda-4\im) }= \nonumber \\
&=0, \mbox{if} ~~  \phi(\lambda)=0, \label{semi}
\end{align}
because for $\phi(\lambda)=0$ the right hand side of (\ref{interm}) is
completely independent of $\delta$.

We next replace (\ref{interm1}) in (\ref{interm2}), which is eventually
substituted in (\ref{semi}) and results in an equation for
$\frac{\partial}{\partial\delta}\log \Lambda_{1}(\lambda)$. Using the relation
between $\Lambda_{1}(\lambda)$ and $\Lambda^{[1]}(\lambda)$ (\ref{diff-norm}),
we can re-write the equation (\ref{semi}) in terms of the $\omega$-function
(\ref{omega}) or its simply related $\Omega$-function given by \cite{KNS013},
\eq
\Omega^{[3]}(\lambda,\mu):=2\im \frac{\omega^{[3]}(\lambda,\mu)+\frac{1}{2}}{(\lambda-\mu)^2+4}=-\frac{\partial}{\partial \delta}\log \Lambda^{[1]}(\lambda;\mu))\Big|_{\delta=0} + \frac{2\im}{(\lambda-\mu)^2+4}.
\en
After a long but straightforward calculation, we obtain from (\ref{semi})
the equation (\ref{analogueof74}), where we identified $(\lambda,\mu)$ with  ($\xi_1,\xi_2)$.

We need another equation for the $\omega$-function with arguments differing by
$2\im$. These special arguments appear in the expressions of the scale factors
$N^{[2s]}(\lambda)$ like
$N^{[2]}(\lambda):=\Lambda^{[2]}(\lambda)/\Lambda^{[1]}(\lambda+\im)\Lambda^{[1]}(\lambda-\im)$. The
r.h.s.~of this relation can be expressed as the expectation value of the
projector $\check{P}_1$ onto triplet states in the tensor product of two
spin-$1/2$ objects with respect to the density matrix
\begin{equation}\label{omegaDarstellungDichte1}
D^{[1]}(\lambda,\mu)=\left(\frac14-\frac16\omega(\lambda,\mu)\right)\mbox{Id}+\frac13 \omega(\lambda,\mu) \check{P}_1,
\end{equation}
with $\lambda\to\lambda-\im, \mu\to\lambda+\im$ yielding
\begin{equation}
\frac{\Lambda_2(\lambda)}{\Lambda_1(\lambda+\im)\Lambda_1(\lambda-\im)}=\frac{3}{4}+\frac{1}{2}\omega(\lambda-\im,\lambda+\im).\label{N2}
\end{equation}
Hence with $\omega=\omega^{[2]}$ we find
$N^{[2]}(\lambda)=\frac{3}{4}+\frac{1}{2}\omega^{[2]}(\lambda-\im,\lambda+\im)$.
Note that (\ref{omegaDarstellungDichte1}) and (\ref{N2}) hold for all
$\omega=\omega^{[2s]}$ with arbitrary $s$.

For $N^{[3]}(\lambda)$ we find an expression linear in $\omega$ functions by use
of fusion relations like above, however without the $\varphi$ factors stemming
from the insertion of the two auxiliary spin-$1/2$ spaces in
(\ref{monodromy-mod}) which now have to be dropped
\begin{align}
&\Lambda_{2}(\lambda )=\Lambda_{1}(\lambda-\im)\Lambda_{1}(\lambda+ \im)- \phi(\lambda-4\im)\phi(\lambda+4\im),\label{Bfusion2}\\
&\Lambda_{3}(\lambda )=\Lambda_{2}(\lambda-\im)\Lambda_{1}(\lambda+2\im) - \phi(\lambda-3\im)\phi(\lambda+5\im)\Lambda_{1}(\lambda-2\im).\label{Bfusion3}
\end{align}
From (\ref{Bfusion2}) and (\ref{N2}) we obtain
\begin{equation}
\frac{\phi(\lambda-4\im)\phi(\lambda+4\im)}{\Lambda_{1}(\lambda-\im)\Lambda_{1}(\lambda+
  \im)}
= \frac{1}{4}-\frac{1}{2}\omega(\lambda-\im,\lambda+\im).\label{Lam2}
\end{equation}
Dividing (\ref{Bfusion3}) by
$\Lambda_{1}(\lambda-2\im)\Lambda_{1}(\lambda)\Lambda_{1}(\lambda+2\im)$ and
applying (\ref{N2},\ref{Lam2})
\begin{align}
\frac{\Lambda_{3}(\lambda
  )}{\Lambda_{1}(\lambda-2\im)\Lambda_{1}(\lambda)\Lambda_{1}(\lambda+2\im)}&=
\frac{\Lambda_{2}(\lambda-\im)}{\Lambda_{1}(\lambda-2\im)\Lambda_{1}(\lambda)}
-
\frac{\phi(\lambda-3\im)\phi(\lambda+5\im)}{\Lambda_{1}(\lambda)\Lambda_{1}(\lambda+2\im)},\nonumber\\
&=\frac 12\left[1+\omega(\lambda-2\im,\lambda)+\omega(\lambda,\lambda+2\im)\right].
\label{Bfusion4}
\end{align}
Taking the analogue of (\ref{inversion}), i.e.~without $\varphi$ factors, at
$\lambda$ values for which $\phi(\lambda)=0$ we find
\begin{align}
&\frac{\Lambda_{3}(\lambda
  -\im)}{\Lambda_{1}(\lambda-3\im)\Lambda_{1}(\lambda-\im)\Lambda_{1}(\lambda+\im)}\cdot
\frac{\Lambda_{3}(\lambda
  +\im)}{\Lambda_{1}(\lambda-\im)\Lambda_{1}(\lambda+\im)\Lambda_{1}(\lambda+3\im)}
\nonumber\\
&=
\frac{\phi(\lambda-6\im)\phi(\lambda+2\im)}{\Lambda_{1}(\lambda-3\im)\Lambda_{1}(\lambda-\im)}\cdot
\frac{\phi(\lambda-4\im)\phi(\lambda+4\im)}{\Lambda_{1}(\lambda-\im)\Lambda_{1}(\lambda+\im)}\cdot
\frac{\phi(\lambda-2\im)\phi(\lambda+6\im)}{\Lambda_{1}(\lambda+\im)\Lambda_{1}(\lambda+3\im)}.
\label{Bfusion5}
\end{align}
Inserting (\ref{Bfusion4}) on the l.h.s.~and (\ref{Lam2}) on the r.h.s.~we
find after some simple transformations 
\begin{equation}
\omega^{[3]}(\lambda-3\im,\lambda-\im)=-\frac 12 \cdot
\frac{8\,\omega^{[3]}(\lambda-\im,\lambda+\im)+5+6\,\omega^{[3]}(\lambda+\im,\lambda+3\im)}{3+2\,\omega^{[3]}(\lambda+\im,\lambda+3\im)},\label{consist2}
\end{equation}
which will be important for consistently solving the functional equations in
the spin-$3/2$ case. Here we have set $\omega=\omega^{[3]}$. Note that this
equation does not hold for $\omega^{[2]}$.

\section*{\bf Appendix C: Coefficients $\widetilde{\rho}_i(\xi_1,\xi_2)$ }

\begin{footnotesize}
\begin{align*}
&\widetilde{\rho}_1(\xi_1,\xi_2)=\frac{1}{16}(1+\omega(\xi_1^-,\xi_1)+\omega(\xi_1,\xi_1^+))(1+\omega(\xi_2^-,\xi_2)+\omega(\xi_2,\xi_2^+)) \\
&+\frac{1}{60}{\frac{({\xi}^{2}+196)({\xi}^{2}+16)\omega(\xi_1,\xi_2)}{(\xi^2+4)^2}}\\
&-\frac{1}{10}{\frac{(-11{\xi}^{2}+22\im \xi+96)}{{\xi}^{2}(\xi-2\im)^{2}}}[\omega(\xi_1^-,\xi_1)+\omega(\xi_2,\xi_2^+)] \\
&-\frac{1}{10}{\frac{(-11{\xi}^{2}-22\im \xi+96)}{{\xi}^{2}(\xi+2\im)^{2}}}[\omega(\xi_1,\xi_1^+)+\omega(\xi_2^-,\xi_2)]\\
&-\frac{1}{10}\left\{{\frac{(\xi-6\im)(\xi+4\im)}{{\xi}^{2}}}[\omega(\xi_1^-,\xi_2)+\omega(\xi_1,\xi_2^+)]+{\frac{(\xi+6\im)(\xi-4\im)}{{\xi}^{2}}}[\omega(\xi_1^+,\xi_2)+\omega(\xi_1,\xi_2^-)]\right\}\\
&-{\frac{3}{80}}\left\{{\frac{(\xi+2\im)(\xi-4\im)(\xi+14\im)\omega(\xi_1^-,\xi_2^+)}{\xi(\xi-2\im)^2}}+{\frac{(\xi-2\im)(\xi+4\im)(\xi-14\im)\omega(\xi_1^+,\xi_2^-)}{\xi(\xi+2\im)^2}}\right\}\\
&+\frac{1}{25}{\frac{p_1(\xi)\omega(\xi_1,\xi_1^+)\omega(\xi_2^-,\xi_2)+\bar{p}_1(\xi)\omega(\xi_1^-,\xi_1)\omega(\xi_2,\xi_2^+)}{{\xi}^{3}(\xi^2+4)^2}}\\
&+{\frac{3}{25}}{\frac{(11{\xi}^{4}-112{\xi}^{2}+912)}{{\xi}^{2}(\xi^2+4)^2}}[\omega(\xi_1^-,\xi_1)\omega(\xi_2^-,\xi_2)+\omega(\xi_1,\xi_1^+)\omega(\xi_2,\xi_2^+)]\\
&-{\frac{1}{400}}\frac{(11{\xi}^{2}-148)(\xi^2+16)}{{\xi}^{2}(\xi^2+4)}\Big\{\frac{(\xi-6\im)}{(\xi+2\im)}\omega(\xi_1^+,\xi_2^-)[\omega(\xi_1^-,\xi_1)+\omega(\xi_2,\xi_2^+)]\\
&+{\frac{(\xi+6\im)}{(\xi-2\im)}}\omega(\xi_1^-,\xi_2^+)[\omega(\xi_1,\xi_1^+)+\omega(\xi_2^-,\xi_2)]\\
&+{\frac{({\xi}^{2}+36)}{(\xi^2+4)}}\Big[\omega(\xi_1^+,\xi_2^+)\left(\omega(\xi_1^-,\xi_1)+\omega(\xi_2^-,\xi_2)\right)+ \omega(\xi_1^-,\xi_2^-)\left(\omega(\xi_1,\xi_1^+)+\omega(\xi_2,\xi_2^+)\right)\Big] \Big\}\\
&-{\frac{1}{300}}{\frac{(11{\xi}^{2}-148)(\xi^2+36)(\xi-4\im)}{{\xi}^{3}(\xi^2+4)}}[\omega(\xi_1^-,\xi_1)\omega(\xi_1^+,\xi_2)+\omega(\xi_1,\xi_2^-)\omega(\xi_2,\xi_2^+)]\\
&-{\frac{1}{300}}{\frac{(11{\xi}^{2}-148)(\xi^2+36)(\xi+4\im)}{{\xi}^{3}(\xi^2+4)}}[\omega(\xi_1,\xi_1^+)\omega(\xi_1^-,\xi_2)+\omega(\xi_1,\xi_2^+)\omega(\xi_2^-,\xi_2)]\\
&+\frac{1}{6400}{\frac{(3{\xi}^{2}+76)({\xi}^{2}+36)({\xi}^{2}+16)^{2}}{(\xi^2+4)^{2}}}[\omega(\xi_1,\xi_2)\left(\omega(\xi_1^-,\xi_2^-)+\omega(\xi_1^+,\xi_2^+)\right)+\omega(\xi_1^-,\xi_2^-)\omega(\xi_1^+,\xi_2^+)]\\
&-\frac{1}{6400}{\frac{(3{\xi}^{2}+76)({\xi}^{2}+36)({\xi}^{2}+16)}{{\xi}^{2}}}[\omega(\xi_1,\xi_2^+)\omega(\xi_1^+,\xi_2)+\omega(\xi_1,\xi_2^-)\omega(\xi_1^-,\xi_2)]\\
&-\frac{1}{6400}{\frac{(3{\xi}^{2}+76)({\xi}^{2}+16)^{2}\omega(\xi_1^-,\xi_2^+)\omega(\xi_1^+,\xi_2^-)}{(\xi^2+4)}}\\
&+\frac{1}{6400}\frac{(\xi^2+16)}{{\xi}^{3}}\left[{\frac{p_2(\xi)(\xi-6\im)}{(\xi-4\im)}}\omega(\xi_1,\xi_2^+)\omega(\xi_1^-,\xi_2)+\frac{\bar{p}_2(\xi)(\xi+6\im)}{(\xi+4\im)}\omega(\xi_1,\xi_2^-)\omega(\xi_1^+,\xi_2)\right]\\
&-\frac{1}{6400}\frac{(\xi^2+16)}{\xi(\xi^2+4)}\left[{\frac{p_2(\xi)(\xi-4\im)}{(\xi-2\im)}}\omega(\xi_1,\xi_2)\omega(\xi_1^-,\xi_2^+)+{\frac{\bar{p}_2(\xi)(\xi+4\im)}{(\xi+2\im)}}\omega(\xi_1,\xi_2)\omega(\xi_1^+,\xi_2^-)\right]\\
\end{align*}
\begin{align}
&+\frac{1}{6400}{\frac{p_3(\xi)(\xi+6\im)(\xi+2\im)(\xi-4\im)^{2}}{{\xi}^{2}(\xi-2\im)}}[\omega(\xi_1,\xi_2^-)\omega(\xi_1^-,\xi_2^+)+\omega(\xi_1^+,\xi_2)\omega(\xi_1^-,\xi_2^+)]\nonumber\\
&+\frac{1}{6400}{\frac{\bar{p}_3(\xi)(\xi-6\im)(\xi-2\im)(\xi+4\im)^{2}}{{\xi}^{2}(\xi+2\im)}}[\omega(\xi_1,\xi_2^+)\omega(\xi_1^+,\xi_2^-)+\omega(\xi_1^-,\xi_2)\omega(\xi_1^+,\xi_2^-)]\nonumber\\
&-\frac{1}{6400}{\frac{\bar{p}_3(\xi)(\xi^2+36)(\xi^2+16)}{{\xi}^{2}(\xi+2\im)}}[\omega(\xi_1,\xi_2^-)\omega(\xi_1^+,\xi_2^+)+\omega(\xi_1^-,\xi_2^-)\omega(\xi_1^+,\xi_2)]\nonumber\\
&-\frac{1}{6400}{\frac{p_3(\xi)(\xi^2+36)(\xi^2+16)}{{\xi}^{2}(\xi-2\im)}}[\omega(\xi_1,\xi_2^+)\omega(\xi_1^-,\xi_2^-)+\omega(\xi_1^+,\xi_2^+)\omega(\xi_1^-,\xi_2)]\nonumber\\
&+\frac{1}{19200}(3{\xi}^{4}+54{\xi}^{2}+296)\Big\{{\frac{({\xi}^{2}+36)({\xi}^{2}+16)^{2}}{(\xi^2+4)^{2}}}\omega(\xi_1^-,\xi_2^-)\omega(\xi_1,\xi_2)\omega(\xi_1^+,\xi_2^+)\nonumber\\
&+{\frac{(\xi-6\im)(\xi-2\im)(\xi+4\im)^{2}\omega(\xi_1^+,\xi_2^-)\omega(\xi_1,\xi_2^+)\omega(\xi_1^-,\xi_2)}{{\xi}^{2}}}\nonumber\\
&+{\frac{(\xi+6\im)(\xi+2\im)(\xi-4\im)^{2}\omega(\xi_1^-,\xi_2^+)\omega(\xi_1,\xi_2^-)\omega(\xi_1^+,\xi_2)}{{\xi}^{2}}}\nonumber\\
&-{\frac{({\xi}^{2}+36)({\xi}^{2}+16)}{{\xi}^{2}}}[\omega(\xi_1^-,\xi_2^-)\omega(\xi_1,\xi_2^+)\omega(\xi_1^+,\xi_2)+\omega(\xi_1^+,\xi_2^+)\omega(\xi_1,\xi_2^-)\omega(\xi_1^-,\xi_2)]\nonumber\\
&-{\frac{({\xi}^{2}+16)^{2}\omega(\xi_1^-,\xi_2^+)\omega(\xi_1,\xi_2)\omega(\xi_1^+,\xi_2^-)}{(\xi^2+4)}}\Big\}\nonumber\\
&-{\frac{1}{600}}\frac{(11{\xi}^{2}-148)(\xi^2+16)}{{\xi}^{2}(\xi^2+4)}\Big\{\frac{({\xi}^{2}+36)}{(\xi^2+4)}[\omega(\xi_1^-,\xi_1)\omega(\xi_1^+,\xi_2^+)\omega(\xi_2^-,\xi_2)+\omega(\xi_1,\xi_1^+)\omega(\xi_1^-,\xi_2^-)\omega(\xi_2,\xi_2^+)]\nonumber\\
&+\left[\frac{(\xi-6\im)}{(\xi+2\im)}\omega(\xi_1^-,\xi_1)\omega(\xi_1^+,\xi_2^-)\omega(\xi_2,\xi_2^+)+\frac{(\xi+6\im)}{(\xi-2\im)}\omega(\xi_1,\xi_1^+)\omega(\xi_1^-,\xi_2^+)\omega(\xi_2^-,\xi_2)\right]\Big\},
\end{align}
\end{footnotesize}
where $\bar{p}_i(\xi)$ are the complex conjugates of $p_i(\xi)$ which are
defined by
\bear
p_1(\xi)&=&(33{\xi}^{5}-22\im{\xi}^{4}-116{\xi}^{3}+824\im{\xi}^{2}-224 \xi-7104\im), \nonumber\\
p_2(\xi)&=&(3{\xi}^{5}-18\im{\xi}^{4}+4{\xi}^{3}-104\im{\xi}^{2}-1824 \xi-4736\im), \\
p_3(\xi)&=&(3{\xi}^{3}-18\im{\xi}^{2}-4\xi-296\im).\nonumber
\ear

\begin{footnotesize}
\begin{align*}
&\widetilde{\rho}_2(\xi_1,\xi_2)=\frac{1}{16}(1+\omega(\xi_1^-,\xi_1)+\omega(\xi_1,\xi_1^+))(1+\omega(\xi_2^-,\xi_2)+\omega(\xi_2,\xi_2^+))\\
&-\frac{1}{20}{\frac{({\xi}^{2}+100)({\xi}^{2}+16)\omega(\xi_1,\xi_2)}{(\xi^2+4)^2}}\\
&-\frac{3}{5}{\frac{-\im(\xi+6\im)(\xi-4\im)}{{\xi}^{2}(\xi+4\im)}}[\omega(\xi_1^+,\xi_2)+\omega(\xi_1,\xi_2^-)]\\
&-\frac{3}{5}{\frac{ \im(\xi-6\im)(\xi+4\im)}{{\xi}^{2}(\xi-4\im)}}[\omega(\xi_1^-,\xi_2)+\omega(\xi_1,\xi_2^+)]\\
&+\frac{3}{10}{\frac{({\xi}^{2}-2\im \xi-16)({\xi}^{2}-2\im \xi-24)}{{\xi}^{2}(\xi^2+4)(\xi-2\im)(\xi-4\im)}}[\omega(\xi_1^-,\xi_1)+\omega(\xi_2,\xi_2^+)]\\
&+\frac{3}{10}{\frac{({\xi}^{2}+2\im \xi-16)({\xi}^{2}+2\im \xi-24)}{{\xi}^{2}(\xi^2+4)(\xi+2\im)(\xi+4\im)}}[\omega(\xi_1,\xi_1^+)+\omega(\xi_2^-,\xi_2)]\\
&-\frac{3}{40}{\frac{({\xi}^{2}+36)}{(\xi^2+4)}}[\omega(\xi_1^-,\xi_2^-)+\omega(\xi_1^+,\xi_2^+)]\\
&+\frac{3}{80}\left\{\frac{(\xi+4\im)(\xi+2\im)(\xi-10\im)\omega(\xi_1^-,\xi_2^+)}{\xi(\xi-2\im)^2}+\frac{(\xi-4\im)(\xi-2\im)(\xi+10\im)\omega(\xi_1^+,\xi_2^-)}{\xi(\xi+2\im)^2}\right\}\\
&+\frac{1}{64} \left\{ \frac{q_1(\xi)\omega(\xi_1^-,\xi_1)\omega(\xi_2,\xi_2^+)}{{\xi}^{3}(\xi^2+4)^{2}(\xi-4\im)}+\frac{\bar{q}_1(\xi)\omega(\xi_1,\xi_1^+)\omega(\xi_2^-,\xi_2)}{{\xi}^{3}(\xi^2+4)^{2}(\xi+4\im)}\right\}\\
&+\frac{1}{64}\left(-{\frac{3391488}{25}}+{\frac{377856}{25}}{\xi}^{2}-{\frac{4608}{5}}{\xi}^{4}+{\frac{576}{25}}{\xi}^{6}\right)\frac{\left[\omega(\xi_1^-,\xi_1)\omega(\xi_2^-,\xi_2)+\omega(\xi_1,\xi_1^+)\omega(\xi_2,\xi_2^+)\right]}{{\xi}^{2}(\xi^2+16)(\xi^2+4)^{2}}\\
&-\frac{1}{6400}{\frac{({\xi}^{4}+84{\xi}^{2}+1472)({\xi}^{2}+36)({\xi}^{2}+16)}{(\xi^2+4)^2}}[\omega(\xi_1,\xi_2)\left(\omega(\xi_1^-,\xi_2^-)+\omega(\xi_1^+,\xi_2^+)\right)+\omega(\xi_1^-,\xi_2^-)\omega(\xi_1^+,\xi_2^+)]\\
&+\frac{1}{6400}{\frac{({\xi}^{4}+84{\xi}^{2}+1472)({\xi}^{2}+36)}{{\xi}^{2}}}[\omega(\xi_1^-,\xi_2)\omega(\xi_1,\xi_2^-)+\omega(\xi_1^+,\xi_2)\omega(\xi_1,\xi_2^+)]\\
&+\frac{1}{6400}{\frac{({\xi}^{4}+84{\xi}^{2}+1472)({\xi}^{2}+16)\omega(\xi_1^-,\xi_2^+)\omega(\xi_1^+,\xi_2^-)}{(\xi^2+4)}}\\ 
&-\frac{1}{6400}\left\{{\frac{q_2(\xi)(\xi-6\im)(\xi+4\im)^{2}\omega(\xi_1^-,\xi_2)\omega(\xi_1,\xi_2^+)}{{\xi}^{3}(\xi-4\im)}}+{\frac{\bar{q}_2(\xi)(\xi+6\im)(\xi-4\im)^{2}\omega(\xi_1^+,\xi_2)\omega(\xi_1,\xi_2^-)}{{\xi}^{3}(\xi+4\im)}}\right\}\\
&+\frac{1}{6400}\frac{(\xi^2+16)}{\xi(\xi^2+4)}\left\{{\frac{q_2(\xi)(\xi+4\im)\omega(\xi_1^-,\xi_2^+)\omega(\xi_1,\xi_2)}{(\xi-2\im)}}+{\frac{\bar{q}_2(\xi)(\xi-4\im)\omega(\xi_1^+,\xi_2^-)\omega(\xi_1,\xi_2)}{(\xi+2\im)}}\right\}\\
&-\frac{1}{6400}{\frac{q_3(\xi)(\xi+6\im)(\xi-4\im)(\xi+2\im)}{{\xi}^{2}(\xi-2\im)}}[\omega(\xi_1^-,\xi_2^+)\omega(\xi_1,\xi_2^-)+\omega(\xi_1^+,\xi_2)\omega(\xi_1^-,\xi_2^+)]\\
&-\frac{1}{6400}{\frac{\bar{q}_3(\xi)(\xi-6\im)(\xi+4\im)(\xi-2\im)}{{\xi}^{2}(\xi+2\im)}}[\omega(\xi_1^+,\xi_2^-)\omega(\xi_1,\xi_2^+)+\omega(\xi_1^-,\xi_2)\omega(\xi_1^+,\xi_2^-)]\\
&+\frac{1}{6400}{\frac{q_3(\xi)(\xi^2+36)(\xi+4\im)}{{\xi}^{2}(\xi-2\im)}}[\omega(\xi_1^-,\xi_2^-)\omega(\xi_1,\xi_2^+)+\omega(\xi_1^-,\xi_2)\omega(\xi_1^+,\xi_2^+)]\\
&+\frac{1}{6400}{\frac{\bar{q}_3(\xi)(\xi^2+36)(\xi-4\im)}{{\xi}^{2}(\xi+2\im)}}[\omega(\xi_1^+,\xi_2^+)\omega(\xi_1,\xi_2^-)+\omega(\xi_1^+,\xi_2)\omega(\xi_1^-,\xi_2^-)]\\
\end{align*}
\begin{align}
&-\frac{3}{400}{\frac{({\xi}^{4}-76{\xi}^{2}+832)(\xi-6\im)}{{\xi}^{2}(\xi^2+4)(\xi+2\im)}}[\omega(\xi_1^-,\xi_1)\omega(\xi_1^+,\xi_2^-)+\omega(\xi_1^+,\xi_2^-)\omega(\xi_2,\xi_2^+)]\nonumber\\
&-\frac{3}{400}{\frac{({\xi}^{4}-76{\xi}^{2}+832)(\xi+6\im)}{{\xi}^{2}(\xi^2+4)(\xi-2\im)}}[\omega(\xi_1,\xi_1^+)\omega(\xi_1^-,\xi_2^+)+\omega(\xi_1^-,\xi_2^+)\omega(\xi_2^-,\xi_2)]\nonumber\\
&-\frac{3}{400}{\frac{({\xi}^{4}-76{\xi}^{2}+832)({\xi}^{2}+36)}{{\xi}^{2}(\xi^2+4)^{2}}}[\omega(\xi_1^+,\xi_2^+)\left(\omega(\xi_2^-,\xi_2)+\omega(\xi_1^-,\xi_1)\right)+\omega(\xi_1^-,\xi_2^-)\left(\omega(\xi_2,\xi_2^+)+\omega(\xi_1,\xi_1^+)\right)]\nonumber\\
&-\frac{1}{100}{\frac{({\xi}^{4}-76{\xi}^{2}+832)(\xi^2+36)}{{\xi}^{3}(\xi^2+4)(\xi+4\im)}}[\omega(\xi_1^-,\xi_1)\omega(\xi_1^+,\xi_2)+\omega(\xi_1,\xi_2^-)\omega(\xi_2,\xi_2^+)]\nonumber\\
&-\frac{1}{100}{\frac{({\xi}^{4}-76{\xi}^{2}+832)(\xi^2+36)}{{\xi}^{3}(\xi^2+4)(\xi-4\im)}}[\omega(\xi_1,\xi_1^+)\omega(\xi_1^-,\xi_2)+\omega(\xi_1,\xi_2^+)\omega(\xi_2^-,\xi_2)]\nonumber\\
&-\frac{1}{200}\frac{({\xi}^{4}-76{\xi}^{2}+832)}{{\xi}^{2}(\xi^2+4)}\Big\{\frac{({\xi}^{2}+36)}{(\xi^2+4)}[\omega(\xi_1^-,\xi_1)\omega(\xi_1^+,\xi_2^+)\omega(\xi_2^-,\xi_2)+\omega(\xi_1,\xi_1^+)\omega(\xi_1^-,\xi_2^-)\omega(\xi_2,\xi_2^+)]\nonumber\\
&+{\frac{(\xi-6\im)}{(\xi+2\im)}}\omega(\xi_1^-,\xi_1)\omega(\xi_1^+,\xi_2^-)\omega(\xi_2,\xi_2^+)+{\frac{(\xi+6\im)}{(\xi-2\im)}}\omega(\xi_1,\xi_1^+)\omega(\xi_1^-,\xi_2^+)\omega(\xi_2^-,\xi_2)\Big\}\nonumber\\
&+\frac{1}{19200}({\xi}^{4}+34{\xi}^{2}+312)\Big\{{\frac{({\xi}^{2}+16)^{2}}{(\xi^2+4)}}\omega(\xi_1^-,\xi_2^+)\omega(\xi_1,\xi_2)\omega(\xi_1^+,\xi_2^-) \nonumber\\
&+{\frac{({\xi}^{2}+36)({\xi}^{2}+16)}{{\xi}^{2}}}[\omega(\xi_1^+,\xi_2)\omega(\xi_1,\xi_2^+)\omega(\xi_1^-,\xi_2^-)+\omega(\xi_1^-,\xi_2)\omega(\xi_1,\xi_2^-)\omega(\xi_1^+,\xi_2^+)]\nonumber\\
&-{\frac{(\xi-6\im)(\xi+4\im)^{2}(\xi-2\im)\omega(\xi_1^-,\xi_2)\omega(\xi_1,\xi_2^+)\omega(\xi_1^+,\xi_2^-)}{{\xi}^{2}}}\nonumber\\
&-{\frac{(\xi+6\im)(\xi-4\im)^{2}(\xi+2\im)\omega(\xi_1^+,\xi_2)\omega(\xi_1,\xi_2^-)\omega(\xi_1^-,\xi_2^+)}{{\xi}^{2}}}\nonumber\\
&-{\frac{({\xi}^{2}+36)({\xi}^{2}+16)^{2}\omega(\xi_1^-,\xi_2^-)\omega(\xi_1,\xi_2)\omega(\xi_1^+,\xi_2^+)}{(\xi^2+4)^{2}}}\Big\},
\end{align}
\end{footnotesize}
where $\bar{q}_i(\xi)$ are the complex conjugates of $q_i(\xi)$ which are
defined by
\begin{footnotesize}
\bear
q_1(\xi)&=&(-{\frac{1916928}{25}}-{\frac{86016}{5}}\im \xi+{\frac{362496}{25}}{\xi}^{2}+{\frac{13824}{5}}\im{\xi}^{3}-{\frac{5376}{5}} 
{\xi}^{4}-{\frac{384}{5}}\im{\xi}^{5}+{\frac{576}{25}}{\xi}^{6}),\nonumber\\
q_2(\xi)&=&({\xi}^{5}-14\im{\xi}^{4}-36{\xi}^{3}+104\im{\xi}^{2}-288 \xi+4992\im), \\
q_3(\xi)&=&({\xi}^{4}-10\im{\xi}^{3}-36{\xi}^{2}-200\im \xi-1248).\nonumber
\ear
\end{footnotesize}
Due to the normalization property of the density matrix $\tr
D_2^{[2s]}(\xi_1,\xi_2)=1$, we can write the fourth coefficient in
terms of the previous ones. Therefore we have
$\rho_3^{[3]}(\xi_1,\xi_2)=\frac{1}{7}\left(1-\rho_0^{[3]}(\xi_1,\xi_2)-3\rho_1^{[3]}(\xi_1,\xi_2)-5\rho_2^{[3]}(\xi_1,\xi_2) \right)$.

\end{document}